\documentclass[
 amsmath,amssymb,
 aps, prl, twocolumn, superscriptaddress
]{revtex4-2}

\usepackage{mathtools}
\usepackage{amsmath,amssymb}
\usepackage{graphics,graphicx,color, calc}
\usepackage{xcolor}
\usepackage{dcolumn}
\usepackage[english]{babel}
\usepackage{mathrsfs}
\usepackage[mathcal]{eucal}
\usepackage{bm}
\usepackage{lipsum}
\usepackage{xr}
\usepackage{stackengine}

\newlength\raisedepth
\newcommand{\mb}[1]{\mathbf{#1}}

\providecommand{\abs}[1]{\left\lvert #1 \right\rvert}

\newcommand{\eps}{\varepsilon}
\def\[{\left[}
\def\]{\right]}
\def\({\left(}
\def\){\right)}

\newcommand{\jfi}
{\affiliation{James Franck Institute, The University of Chicago, Chicago, IL 60637}}
\newcommand{\phy}
{\affiliation{Department of Physics, The University of Chicago, Chicago, IL 60637}}
\newcommand{\kadanoff}
{\affiliation{Kadanoff Center for Theoretical Physics, The University of Chicago, Chicago, Illinois 60637, USA}}
\newcommand{\princeton}
{\affiliation{Princeton Center for Theoretical Science, Princeton University, Princeton, NJ 08544}}

\begin{document}
\title{Odd Elasticity in Driven Granular Matter}

\author{Rosalind Huang}%
\thanks{These authors contributed equally to this work.}
\jfi
\phy

\author{Rituparno Mandal}%
\thanks{These authors contributed equally to this work.}
\jfi

\author{Colin Scheibner}%
\princeton

\author{Vincenzo Vitelli}%
\email{vitelli@uchicago.edu}
\jfi
\phy
\kadanoff

\begin{abstract}
Odd elasticity describes the unusual elastic response of solids whose stress-strain relationship is not compatible with an elastic potential. Here, we present a study of odd elasticity in a driven granular matter system composed of grains with ratchet-like interparticle friction and activated by oscillatory shear. We find that the system permits a time-averaged elasticity theory featuring nonzero odd elastic coefficients. These coefficients are explicitly measured using molecular dynamics simulations and can be predicted and tuned from microscopics. In the presence of disorder, our driven granular material displays distinctive properties ranging from self-healing grain boundaries in polycrystalline systems to chiral plastic vortices and force chain deflection in amorphous packings. Beyond granular matter, our work motivates the search for microscopic transduction mechanisms that convert periodic nonuniform drive into uniform elastic properties of active solids.

\end{abstract}

\maketitle

The theory of elasticity is built upon the assumption of a well-defined elastic potential energy \cite{landau_theory_2009}. However, in non-equilibrium solids ranging from dense living matter to driven metamaterials, this assumption is often broken by non-conservative microscopic interactions. Odd elasticity refers to the additional elastic moduli that are ordinarily forbidden by energy conservation~\cite{scheibner_odd_2020,fruchart_odd_2023, scheibner_non-hermitian_2020, yasuda_odd_2021, lin_onsagers_2023, zhou_non-hermitian_2020,  cohen_odd_2023-1, ishimoto_self-organized_2022,
cheng_odd_2021, fossati_odd_2022, surowka_odd_2023, chen_odd_2022, ishimoto_odd_2023-1, kole_dynamics_2023} and can in principle arise in these systems. Odd elasticity has been engineered into mechanical metamaterials~\cite{brandenbourger_limit_2022, chen_realization_2021} and has been suggested as an explanation for experimental observations~ \cite{bililign_motile_2022, tan_odd_2022}. Here, we study a granular matter system with chiral friction that is driven by oscillatory shear, and we find that odd elasticity emerges as a time-averaged continuum description. Compared to systems containing active spinners~\cite{bililign_motile_2022, tan_odd_2022} or point-like transverse interactions~\cite{braverman_topological_2021}, our odd granular solid yields distinct phenomenology ranging from
dynamics of defects and grain boundaries, chiral plastic vortices, to the deflection of force chains.

\begin{figure*}[]
\centering
\hspace{-2em}\includegraphics[width = 1.9\columnwidth]{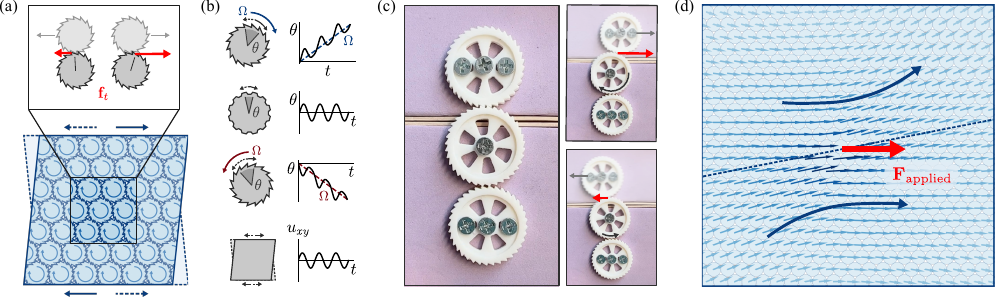}
\caption{A driven granular material with chiral friction. (a) Schematic of the system composed of grains with chiral friction. Top: gray arrows indicate velocities of top particle, red arrows show frictional force on the lower particle. Bottom: blue arrows indicate the oscillatory driving shear. (b) Angular displacements over several uniform oscillatory shear cycles. (c) A simple demonstration of chiral friction and emergent spinning. As the top layer oscillates, the ratcheting effect causes the center particle to rotate clockwise. Note that the driving amplitude is exaggerated compared to the simulations, where we keep the amplitude small to maintain unbroken contacts. (d) Deformation field of the material under a point force.}
\label{fig1}
\end{figure*}

\textit{Model}\textemdash Consider a two-dimensional granular solid \cite{reynolds_lvii_1885, bagnold_physics_1971, de_gennes_granular_1999, jaeger_granular_1996, brown_principles_1970, jaeger_physics_1992, behringer_scientist_1994, bideau_disorder_1993, mehta_dynamics_1994,behringer_physics_2019} composed of particles whose positions $\mb{r}_i$ and angular displacements $\mb{\theta}_i$ follow the underdamped equations \cite{cundall_discrete_1979, silbert_jamming_2010, luding_cohesive_2008, mari_shear_2014}
\begin{align}
	m_i \ddot{\mathbf{r}}_i &= \mathbf{F}_{i}^{\text{int}} - \zeta\dot{\mathbf{r}}_i \label{F_i}\\ 	
	I_i \ddot{\theta}_i &= \tau_i^{\text{int}} - \zeta^{r}\dot{\theta_i}  \label{tau_i},
\end{align}
where for the $i^{\text{th}}$ particle $\mathbf{F}_{i}^{\text{int}}$ and $\boldsymbol{\tau}_i^{\text{int}}$ are total interaction forces and torques and $\zeta\dot{\mathbf{r}}_i$ and $\zeta^{r}\dot{\theta_i}$ describe substrate resistance. Interaction forces $\mathbf{F}_{i}^{\text{int}} = \sum_j (\mathbf{f}_{n_{ij}} + \mathbf{f}_{t_{ij}})$ involve a soft, repulsive harmonic normal force $\mathbf{f}_{n_{ij}}$ and a tangential force $\mathbf{f}_{t_{ij}}$ representing frictional forces~\cite{SM}. We consider a chiral modification of Coulomb's law of contact friction in which the sliding threshold depends on the direction of the force~\cite{zhao_odd_2022, SM}, namely 
\begin{gather}
	\abs{\mathbf{f}_{t_{ij}}} \leq \mu_\pm \abs{\mathbf{f}_{n_{ij}}},\quad \, \mu_\pm = \mu\,(1\pm\eps). \label{chifr}
\end{gather}
As illustrated in Fig.~\ref{fig1}(a), Eq.~(\ref{chifr}) features a coefficient of friction $\mu$ scaled by a factor of $(1+\eps)$ or $(1-\eps)$ depending on whether grains slide along or against their ratchet ``teeth". The degree of chirality is set by $\eps$.

To inject energy, we subject the granular medium to a symmetric oscillatory shear uniform throughout (implemented via the SLLOD algorithm), mimicking a driven viscous suspension of particles \cite{SM,evans_statistical_2008}. The oscillatory shear produces ratchet-like motion of the grains as they interact with their neighbors. Over many cycles, the grains acquire an average spinning speed $\Omega$ that depends on their degree of chirality. (Fig. \ref{fig1}(a)-(b) and Movie S1). These interactions influence the system's collective response to perturbations. When the system in Fig. \ref{fig1}(d) is subjected to a horizontal point force, it experiences a smooth deflection of the particles' time-averaged displacement. The displacement field breaks the up-down symmetry of the lattice and of the horizontal point force, indicating that the effects of chirality and driving have percolated to the level of the continuum description.

\begin{figure}[]
\centering
\hspace{-1em}\includegraphics[width = .9\columnwidth]{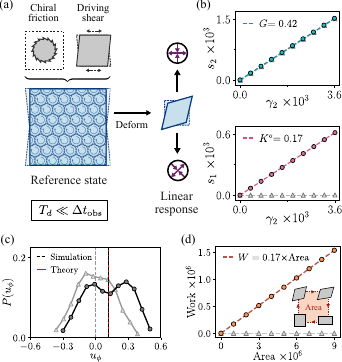}
\caption{Measurement of odd moduli. (a) Schematic of measurement protocol. In (b)-(d), circles represent the driven chiral system, and grey triangles the passive achiral one. (b) The moduli $G$ and $K^o$ are measured as slopes of the stress-strain plots. (c) Distributions of $u_\phi$, the angle of the strain field relative to the point force in Fig.~\ref{fig1}d. The vertical lines indicate averages. (d) Work rectified in a strain cycle as a function of the area it encloses in strain space.}
\label{fig2}
\end{figure}

\textit{Odd elastic moduli}\textemdash Over each driving period $T_d = 2\pi/\omega_d$, we find that the system has a well-defined reference state, where particles sit at their time-averaged positions and spin at the emergent speed $\Omega$ (Fig. \ref{fig2}(a)). We study the system's response to quasistatic deformations \cite{maloney_amorphous_2006-1} that occur on time scales much longer than $T_d$. This form of elasticity captures the relation between \textit{time-averaged} stresses $\sigma_{ij}$ and strains $u_{kl}$. In the linear regime, one expects the relation $\sigma_{ij} = C_{ijkl} u_{kl}$, where $C_{ijkl}$ is the elasticity tensor~\cite{landau_theory_2009}. Of special importance here are the elastic moduli relating shear stresses to shear strains (see~\cite{SM} for general expression),
\begin{align}
	\underset{\sigma_\alpha}{\begin{pmatrix}
		\raisebox{-\raisedepth}{\includegraphics[height=.9em]{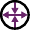}}\\
		\raisebox{-\raisedepth}{\includegraphics[height=.9em]{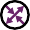}}
	\end{pmatrix}}
	=
	\underset{C_{\alpha\beta}}{\begin{pmatrix}
		{\color[rgb]{0.295347,0.736798,0.752687}G_{1}} &  {\color[rgb]{0.599997,0.600015,0.600005}K^e}{\color[rgb]{0.851358,0.364682,0.482766}+K^o}  \\
		{\color[rgb]{0.599997,0.600015,0.600005}K^e}{\color[rgb]{0.851358,0.364682,0.482766}-K^o}  & {\color[rgb]{0.295347,0.736798,0.752687}G_{2}}
	\end{pmatrix}}
	\underset{e_\beta}{\begin{pmatrix}
		\raisebox{-\raisedepth}{\includegraphics[height=.75em]{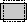}}\\
		\raisebox{-\raisedepth}{\includegraphics[height=.9em]{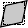}}
	\end{pmatrix}}.
 \label{constitutive}
\end{align}
The stresses and strains are decomposed into components represented pictorially in Eq. (\ref{constitutive}) \cite{SM,scheibner_odd_2020, fruchart_odd_2023}. Here $G_1$ and $G_2$ denote the conservative shear moduli, and $K^o$ is the odd shear modulus. Crucially, the oscillatory shear establishes anistropy in the system by breaking rotational symmetry, which can appear as $G_1\neq G_2$ and $K^e\neq 0$ in Eq. (\ref{constitutive}).

The stress-strain plots of Fig. \ref{fig2}(b) present the measurement results for the driven system with chirality $\eps = 0.6$. The material retains the shear modulus $G$ of standard granular solids, but we find a linear, off-diagonal response of shear stress 1 (\raisebox{-\raisedepth}{\includegraphics[height=.9em]{figures/icons_s2.pdf}}) to the applied pure shear 2 (\raisebox{-\raisedepth}{\includegraphics[height=.9em]{figures/icons_e3.pdf}}) and vice versa, corresponding to an odd modulus $K^o \approx 0.4\,G$. Because the driving amplitude is small and the underlying lattice is isotropic in this measurement, the asymmetry in the moduli due to anisotropy is minor so that $G_1=G_2=G$ and $K^e=0$; however, for stronger driving \cite{SM} or in the presence of defects (discussed later), the effects of anisotropy become more pronounced. Unlike $K^o$, the odd modulus $A$ coupling isotropic compression to torque density \cite{scheibner_odd_2020, fruchart_odd_2023}, crucial in the crystals of self-spinning particles studied so far \cite{bililign_motile_2022,tan_odd_2022}, is typically absent in our system where the torque density vanishes under isotropic deformation and zero substrate resistance (see \cite{SM}). 

These moduli are well defined in the sense that they predict deformations in response to loads independent of how they are measured~\cite{solon_pressure_2015}. For example, the average tilting of the point force response in Fig.~\ref{fig1}(d) is accurately predicted by computing the Green's function $G_{ij} (\mb{r})$ with the measured $K^o$ and $A=0$
\cite{SM} (Fig.~\ref{fig2}{c}). In addition, work of magnitude proportional to $K^o$ can be extracted through a quasistatic cycle of shear deformations, rectifying part of the work done in driving the system \cite{scheibner_odd_2020, SM} (Fig.~\ref{fig2}(d)).

\begin{figure}[]
\centering
\hspace{-2em}\includegraphics[width = .9\columnwidth]{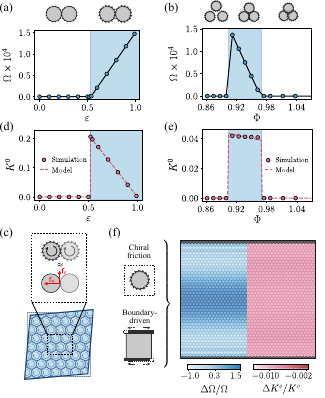}
\caption{Prediction of odd moduli from microscopic interactions. (a)-(b) Emergent spinning speed as functions of (a) chirality ($\eps$) and (b) packing fraction. Shaded regions indicate nonzero $\Omega$. (c) Schematic of the simplified model. The red arrows denote the forces exerted by the right particle on the left. (d) Comparison of simulation to the model's prediction of $K^o$ as a function of (d) $\eps$ and (e) $\Phi$. (f) Spatial variation of spinning speed (left) and odd modulus (right) in the boundary-driven case. Colormaps indicate deviations from mean values. The data for $\Omega$ are presented per layer, while $K^o$ is coarse-grained over a few layers.}
\label{fig3}
\end{figure}

\textit{Predicting and tuning the odd moduli}\textemdash We now examine how the spinning speed and elastic moduli depend on parameters such as the degree of chirality or the packing density. Below a certain threshold in $\eps$, no spinning emerges because the slip force is not reached over the driving cycle (Fig. \ref{fig3}(a)). Above the threshold, the spinning speed $\Omega$ increases linearly with $\eps$. In addition, $\Omega$ is finite only in an intermediate range of packing fractions where particles are in contact but packed loosely enough to slip (Fig. \ref{fig3}(b)). 

To understand the dependence of $K^o$ on system parameters, we propose a simple heuristic model that is useful for predicting the odd shear modulus $K^o$. (See \cite{SM} for discussion on its applicability for other moduli.) In the limit $t_{\text{obs}}\gg T_d$, the grains appear to be in contact and constantly spinning at the emergent speed $\Omega$, and therefore interacting via \textit{sliding} friction that opposes their spinning direction. 
Under shear deformations, the sliding friction forces change according to $\mu (1- \eps)$ times the normal forces. We find that under shear transformations in particular, the time-averaged tangential forces change along with the sliding forces, which suggests that
\begin{align}
	K^o = \mu\,(1-\eps) \, G
	\label{Ko_prd}
\end{align}
whenever $\Omega\neq 0$, and $K^o = 0$ when $\Omega= 0$  \cite{SM}. The prediction in Eq.~(\ref{Ko_prd}) compares favorably with numerical results shown in Fig. \ref{fig3}(d)-(e). Crucially, Eq.~(\ref{Ko_prd}) suggests that the magnitude of $K^o$ does \textit{not} depend on $\Omega$. In our system where the odd modulus arises directly from frictional forces, $K^o$ inherits the Coulomb's law property of sliding friction: the frictional force does not depend on the relative speed of contact points. The spinning serves to establish a consistent orientation of the tangential force, but thereafter it is the speed-independent sliding friction that determines the odd modulus $K^o$.

\textit{Boundary driven}\textemdash While we have thus far considered oscillatory shear applied uniformly across the medium, it is natural to ask whether a similar effect can be achieved if we drive directly from the boundary.  As the boundary layer oscillates, the motion propagates layer-by-layer into the bulk such that the emergent spinning $\Omega$ varies significantly across layers. In general the system's response to stress will be strongly position-dependent as well, precluding the definition of a coarse-grained modulus \cite{SM}. However, if the boundary is driven sufficiently slowly, the system's response approximates the uniform oscillatory shear in Fig. \ref{fig1}(a). In this case, the modulus $K^o$ becomes more uniform throughout the solid even as $\Omega$ varies (Fig. \ref{fig3}(f) and discussed in \cite{SM}). This is a system that realizes odd elasticity without energy input at the particle scale, and it does so by taking advantage of features particular to granular matter: Coulomb friction and boundary drive.


\begin{figure*}[]
\centering
\hspace{-2em}\includegraphics[width = 1.9\columnwidth]{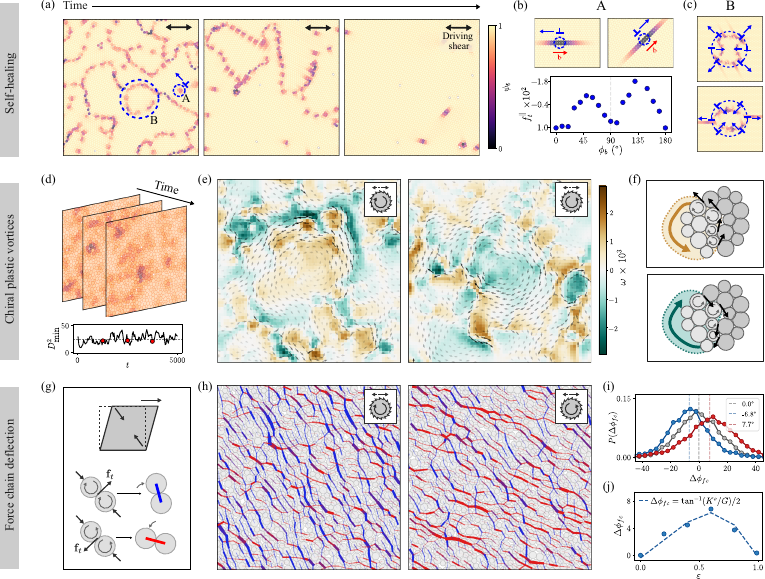}
\caption{Odd elasticity modifies the behavior of systems with disorder. (a) Snapshots of $\psi_6$ shows grain boundary merging over time scales of $\sim 10^3 \times T_d$. (b) Defect self-propulsion direction depends on $b_\phi$, the angle between the Burgers vector $\mb{b}$ and the driving shear axis. The plot shows the total force on the sevenfold-coordinated particle projected onto $\mb{b}$, as a function of $b_\phi$. Positive (negative) $f_t^\parallel$ corresponds to motion antiparallel (parallel) to $\mb{b}$. (c) Anisotropic elongation of circular grain boundaries over time. (d) A time series of the $D^2_{\text{min}}$ shows continuous plastic rearrangements. (e) Snapshots of the vorticity $\omega$ for clockwise-spinning ($\eps = 0.6$, left) and counterclockwise-spinning ($\eps = -0.6$, right) grains. (f) Schematic showing the frictional forces (black arrows) on the boundary particles of a patch. (g) Schematic of two-particle interactions under simple shear. (h) Force chains under steady, simple shear. Blue (red) lines are tilted clockwise (counterclockwise) relative to the diagonal. (i) Distribution of force chain deflection angles, with $\Delta \phi_{\text{fc}}=0$ representing the frictionless case. Blue (red) data represent $\eps>0$ ($\eps<0$), and grey data represent the frictionless system. (j) The deflection angle as a function of $K^o/G$ (simulation and theory). }
\label{fig4}
\end{figure*}


\textit{Grain boundary self-healing}\textemdash When grain boundaries are present, we continue to observe emergent spinning and odd response to strain \cite{SM}. In addition, over time scales of $\sim 10^{3} \times T_d$, the grain boundaries are mobile in the absence of any additional loading. It has previously been appreciated that defect proliferation and grain boundary motion due to transverse forces can act to destabilize solids~\cite{bililign_motile_2022}. However, in our system, we observe the opposite: grain boundaries spontaneously rearrange and merge, leaving in the end a nearly pristine crystal (Fig. \ref{fig4}(a) and Movie S2).

To understand the self-healing of grain boundaries, we first isolate single defects and study their dynamical behavior. Isolated defects in driven solids tend to self-propel along their glide planes (the axis parallel to their Burgers vector)~\cite{braverman_topological_2021, bililign_motile_2022}. In our system, the vanishing torque density \cite{SM} implies that defect propulsion cannot be understood from Peach-Koehler forces, but require microscopic considerations. Additionally, the anisotropy created by the oscillatory shear axis strongly influences the defect propulsion direction. When the glide plane forms an angle of around $0^\circ$ or $90^\circ$ with driving shear, the dislocation travels antiparallel to its Burgers vector, while at $45^\circ$ and $135^\circ$ it travels parallel to the Burgers vector. 

This anisotropic response, summarized in Fig.~\ref{fig4}(b), generically suggests a mechanism for grain boundary healing. 
Consider a circular grain boundary formed by defects with radially oriented Burgers vectors (Fig. \ref{fig4}(c)). The angle dependence of defect motion ensures that the circular grain boundary always elongates along one axis and compresses along another. As a result, the circle shrinks regardless of whether the Burgers vectors face inwards or outwards. An isotropic odd solid, in contrast, would see uniform expansion of the circle in one case and uniform contraction in the other \cite{bililign_motile_2022}. We also note that the triangular lattice is stable in our system approximately when $\mu \, (1-\eps) \leq 1$, i.e. in the regime where normal forces dominate \cite{SM}. 

Combining these results, a picture of grain boundary self-healing emerges. Tangential forces enable defects to self-propel, and the anisotropy enforced by driving shear causes grain boundaries to elongate and merge. Once annihilated, the dislocations do not reappear from within a crystalline region due to the stability of the crystal.

\textit{Chiral plastic vortices}\textemdash We now investigate amorphous packings of chiral particles, using a bidisperse mixture with radii 1 and 1.4 which inhibits crystallization entirely \cite{speedy_glass_1999, perera_stability_1999, ohern_random_2002,chacko_slow_2019}. We drive the system through oscillatory shear, and the particles acquire an emergent spinning speed as before \cite{SM}. Subjected to a sufficient amplitude of driving oscillations, we find that frictional amorphous systems (both achiral or chiral) enter a steady state characterized by continuous plastic deformations \cite{maloney_amorphous_2006-1, chattoraj_elastic_2013, chikkadi_long-range_2011, radjai_turbulentlike_2002, sun_turbulent-like_2022, sun_turbulent-like_2022-1, goldenberg_particle_2007} over time scales $\sim 10^{3} \times T_d$ in our numerical simulations (Fig. \ref{fig4}(d)). 

Over these slow time scales, the displacement field fragments into transient vortical patches with definite handedness. We capture this through the vorticity $\omega = \nabla \times \mb{v}$, where $\mb{v}$ is the coarse-grained, timed-averaged velocity of the grains (Fig. \ref{fig4}(e) and Movie S3). Upon closer inspection, we observe that the deformation vortices follow the \textit{opposite} handedness to the spinning direction $\Omega$, in contrast to whorls observed in systems of self-spinning particles  \cite{bililign_motile_2022}. 
We can rationalize our observations by considering the tangential forces (Fig. \ref{fig4}(f)). Density fluctuations can cause a region to momentarily form a rigid body-like patch, whose boundary particles spin at the average speed $\Omega$. As their spin is resisted by friction against outside neighbors, a torque opposing the spinning direction of each grain is imparted on the patch.

\textit{Dynamic force chain deflection}\textemdash When a granular material without odd elasticity is deformed, force chains \cite{cates_jamming_1998, radjai_force_1996, drescher_photoelastic_1972, liu_force_1995, majmudar_contact_2005, makse_packing_2000, brodu_spanning_2015, brujic_3d_2003, desmond_experimental_2013, katgert_jamming_2010, mandal_extreme_2020, delarue_self-driven_2016} form on average along an axis of compression. In our amorphous system, we can measure a well-defined odd elastic response in the amorphous system over a time scale $\sim 10^2\times T_d$ \cite{SM}. The force chains in this odd amorphous system deviate from the compression axis when subjected to steady simple shear (Fig. \ref{fig4}(h)-(i)  and Movie S4). Frictional forces between particles that resist their spinning cause bonds to tilt as they are compressed, and the direction of tilt is determined by the spinning handedness (Fig. \ref{fig4}(g)). This microscopic picture can be systematically captured by the odd modulus $K^o$. According to the stress-strain relation Eq. (\ref{constitutive}), the stress responds to the pure shear (\raisebox{-\raisedepth}{\includegraphics[height=.9em]{figures/icons_e3.pdf}}) component of the applied simple shear via a combination of the two shear stresses $\sigma  \sim  {\color[rgb]{0.295347,0.736798,0.752687}G}\,\raisebox{-\raisedepth}{\includegraphics[height=.9em]{figures/icons_s3.pdf}} + {\color[rgb]{0.851358,0.364682,0.482766}K^o}\,\raisebox{-\raisedepth}{\includegraphics[height=.9em]{figures/icons_s2.pdf}}$. Geometrically, this corresponds to a stress field that is rotated by a clockwise angle of $\sim\arctan(K^o/G)/2$ to the diagonal (see ref. \cite{cohen_odd_2023-1} for analogous calculations for displacements). By comparing this prediction to the force chain deflection angle (Fig. \ref{fig4}(j)), we see that the force chains track the rotation (under the influence of $K^o$) of the system's stress response. Thus, force chains provide a window to observe odd elasticity and its distinctive phenomenological signatures in disordered granular matter.


\bibliography{references}

\end{document}


\title{Supplementary Materials}
\maketitle
\onecolumngrid
\tableofcontents

\vfill
\pagebreak

\section{Simulation procedures}


\subsection{Contact friction}
\label{1_contact}
To model the 2D granular matter system, we perform molecular dynamics (MD) simulations of frictional particles. We simulate $N$ granular discs with masses $m_i$ and radii $R_i$, and positions $\mb{r}_i$ and angular displacements $\mb{\theta_i}$ following the underdamped equations of motion Eqs. (1)-(2). For the interaction forces, we adopt the spring-dashpot model of contact forces \cite{cundall_discrete_1979, silbert_granular_2001, luding_cohesive_2008, mari_shear_2014}. When two particles $i$ and $j$ are in contact, their interaction forces are given by
\begin{align}
\mathbf{f}_{n_{ij}} & =  k_n \delta_{ij} \hat{\mb{r}}_{ij} - \zeta_n \mathbf{v}_{n_{ij}} 	\label{fn_ij}\\
\mathbf{f}_{t_{ij}} & =  -k_t \hat{\mb{s}}_{ij} - \zeta_t \mathbf{v}_{t_{ij}}.   	\label{ft_ij}
\end{align}
The terms $\mb{r}_{ij} = \mb{r}_i - \mb{r}_j$ are relative particle coordinates, and $\mb{v}_{ij}$ are relative velocities of the point of contact (with subscripts $n$ and $t$ denoting normal and tangential projections). The term $\delta_{ij} = R_i+R_j - r_{ij}$ is the normal overlap, and $\mb{s}_{ij}$ keeps track of tangential displacements throughout the duration of contact to model static friction. The $\mb{s}_{ij}$ term starts from $0$ when the contact between particles $i$ and $j$ is first established, and is set to $0$ when the contact is broken. Thereafter, it is integrated numerically via $\dot{\mb{s}}_{ij} = \mathbf{v}_{t_{ij}}-(\mathbf{v}_{t_{ij}}\cdot\mb{s}_{ij}\mb{r}_{ij})/r_{ij}^2$. This means that $\mb{s}_{ij}$ cannot be determined from an instantaneous configuration of the system. The tangential force due to friction, which involves the $\mb{s}_{ij}$ term, is therefore history-dependent~\cite{lemaitre_stress_2021, chattoraj_oscillatory_2019}. The normal and tangential spring constants are $k_n$ and $k_t$, and likewise $\zeta_n$ and $\zeta_t$ are the normal and tangential damping coefficients. The Coulomb condition
\begin{equation}
    \abs{\mathbf{f}_{t_{ij}}} \leq \mu \abs{\mathbf{f}_{n_{ij}}}\label{Coulomb}
\end{equation}
is implemented by truncating the magnitude of $\mathbf{f}_{t_{ij}}$ to satisfy the inequality. In other words, $\mathbf{f}_{t_{ij}}$ is given by Eq. (\ref{ft_ij}) when the inequality is satisfied, and truncated to $\mathbf{f}_{t_{ij}} = \mu \abs{\mathbf{f}_{n_{ij}}} \hat{\boldsymbol{\phi}}_{ij}$ when the inequality is exceeded where $\hat{\boldsymbol{\phi}}_{ij} = \hat{\mb{z}}\times \hat{\mb{r}}_{ij}$. Note that in the main text, $\mu = \mu_\pm$ is modified according to the sign of $\mathbf{v}_{ij}\times \hat{\mb{r}}_{ij}$ to implement chiral friction (see main text Eq. (3). The total interaction force and torque on the $i^{\text{th}}$ particle are
\begin{align}
	\mathbf{F}_{i}^{\text{int}} &= \sum_{j\ne i} 
 (\mathbf{f}_{n_{ij}} + \mathbf{f}_{t_{ij}}) \label{F_tot}\\
\tau_i^{\text{int}} &= -\sum_{j\ne i}\left( \frac{R_i}{R_i+R_j} \mb{r}_{ij} \times \mathbf{f}_{t_{ij}} \right) \cdot \hat{\mb{z}} \label{tau_tot}.
\end{align}

Numerical integration is performed using the velocity-Verlet algorithm \cite{allen_computer_2017} with time steps of $10^{-2}$, and we use system sizes between $N = 1000$ and $4000$. In Table \ref{params} we present typical values of the physical parameters in our simulation. In addition, particle radii are set to 1 for the monodispserse system and set to 1 and 1.4 for the amorphous system, and masses are computed by assuming densities are 1. The drag coefficients $\zeta$ and $\zeta^r$ are set to 0 unless otherwise specified. We chose coefficient of friction $\mu=1$ to represent the higher end of typical frictional materials, whose values are between $\mu = 0.3$ and 1 \cite{malyshev_8_2014}. In accordance with Eqs. (\ref{ft_ij}) and (\ref{Coulomb}), sliding begins when a contact point has undergone a relative displacement of approximately $s_{ij} \sim k_n \delta_{ij} / k_t$. We selected values of the ratio $k_t/k_n$ between $10^{1}$ and $10^{2}$ to keep $s_{ij}$ than around $0.01$ of the particle radii.\\

\begin{table}[h!]
\begin{tabular}{@{}lll@{}}
\toprule
Parameter  \hspace{10em}                    & Symbol    & Values \\ \midrule
Fluid/substrate resistance (translation) & $\zeta$   & 0 (unless stated)       \\
Fluid/substrate resistance (rotation)    & $\zeta^r$ & 0 (unless stated)       \\
Friction coefficient           & $\mu$      & 1              \\ 
Normal spring                  & $k_n$     & 1              \\
Normal damping coefficient     & $\zeta_n$ & 1              \\
Tangential spring              & $k_t$     & 10-100         \\
Tangential damping coefficient & $\zeta_t$ & 0.5            \\
\bottomrule
\end{tabular}
\caption{Table of parameters.}
\label{params}
\end{table}

\subsection{Boundary conditions and shear}
For compatibility with shear, we implement Lees-Edwards periodic boundary conditions throughout our simulations \cite{lees_computer_1972}. There are three methods of shear that we use, depending on the physical context. \textit{Quasistatic shear} is used only for measuring the linear response, and is detailed in Section \ref{protocol}. Two other methods, \textit{SLLOD} and \textit{boundary-driving}, pertain to the means of injecting energy into the granular solid. (See Supplementary Movies for a depiction of the two driving mechanisms.)\\

The SLLOD algorithm \cite{evans_nonlinear-response_1984, evans_statistical_2008} imposes a homogeneous shear strain rate profile $\dot{\mb{u}}(t)$ by modifying the equation of motion Eq. (1) as follows
\begin{align}
	m_i \ddot{\mathbf{r_i}} &= \mathbf{F}_{i}^{\text{int}} - \zeta\left(\dot{\mathbf{r}}_i - \mathbf{r}_i\cdot (\nabla\mb{\dot{u}}(t))^\intercal\right) 	+ m_i \mathbf{r}_i\cdot (\nabla\mb{\ddot{u}}(t)) ^\intercal.\label{sllod}
\end{align}
The second term is now the dissipation relative to the local streaming velocity, and the third term is the force field introduced by SLLOD to sustain the desired velocity profile. For our purposes, we implement simple oscillatory shear in the $x$ direction with strain amplitude $\gamma_d$:
\begin{gather}
    \nabla\mb{u}(t) = 
    \begin{pmatrix}
        0 & \gamma_d\sin(\omega_dt)\\
        0 & 0
    \end{pmatrix},
\end{gather}
leading to the equation of motion
\begin{gather}
    m_i \ddot{\mathbf{r_i}} = \mathbf{F}_{i}^{\text{int}} - \zeta\left(\dot{\mathbf{r}}_i - y_i\cdot \gamma_d \omega_d\cos(\omega_dt) \mb{\hat{x}}\right) 	- m_iy_i\cdot \gamma_d \omega_d^2\sin(\omega_dt) \mb{\hat{x}}. \label{sllod_osc}
\end{gather}
When the ratio $m_i \omega_d / \zeta \ll 1$, the third term on the right becomes insignificant relative to the second term. In this limit, SLLOD can be thought of as embedding the particles in a viscous fluid that is undergoing a shear rate $\dee_y \dot{u}_x(t) = \gamma_d \omega_d\cos(\omega_dt)$. While the SLLOD algorithm is effective in implementing spatially uniform deformations, it involves introducing a fictitious force field as in Eqs. (\ref{sllod}) and (\ref{sllod_osc}) that can be difficult to achieve in experiments. \\

Thus, we also consider boundary-driven shear. In this case, the top and bottom layers of particles are designated as ``walls", and they are forced to oscillate: $\mb{r}_i(t) = \mb{r}_i(0) + y_i \cdot \gamma_d \sin(\omega_dt)\mb{\hat{x}}$. All other particles follow the original equations of motion Eq. (1).

\vfill
\pagebreak
\section{Linear elasticity}

\subsection{Elastic modulus tensor}
Throughout the main text, we adopt the conventions of Refs. \cite{scheibner_odd_2020, fruchart_odd_2023, braverman_topological_2021} for describing stresses, strains, and elastic moduli. Through a change of basis, the rank-2 tensors for stress $\sigma_{ij}$ and strain $u_{ij} = \dee_j u_i$ become 4-dimensional vectors $\sigma_\alpha$ and $e_\alpha$:
\begin{table}[h!]
\begin{tabular}{@{}lll@{}}
\toprule
\multicolumn{1}{c}{Stress}                & \multicolumn{1}{c}{Strain} & \multicolumn{1}{c}{Interpretation} \\ \midrule
$\sigma_0 = \raisebox{-\raisedepth}{\includegraphics[height=.9em]{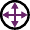}} = (\sigma_{xx}+\sigma_{yy})/2 $ \hspace{1em} & $e_0 = \raisebox{-\raisedepth}{\includegraphics[height=.9em]{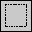}} = u_{xx} + u_{yy}$ \hspace{1em}    & Isotropic dilation \\
$\sigma_1 = \raisebox{-\raisedepth}{\includegraphics[height=.9em]{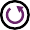}} = (-\sigma_{xy}+\sigma_{yx})/2 $ \hspace{1em} & $e_1 = \raisebox{-\raisedepth}{\includegraphics[height=.9em]{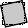}} = u_{xy} - u_{yx}$ \hspace{1em} & Rotation                                    \\
$\sigma_2 = \raisebox{-\raisedepth}{\includegraphics[height=.9em]{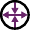}} = (\sigma_{xx}-\sigma_{yy})/2 $ \hspace{1em} & $e_2 = \raisebox{-\raisedepth}{\includegraphics[height=.8em]{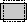}} = u_{xx} - u_{yy}$ \hspace{1em} & Pure shear 1                                    \\
$\sigma_3 = \raisebox{-\raisedepth}{\includegraphics[height=.9em]{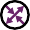}} = (\sigma_{xy}+\sigma_{yx})/2 $ \hspace{1em} & $e_3 = \raisebox{-\raisedepth}{\includegraphics[height=.9em]{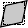}} = u_{xy} + u_{yx}$ \hspace{1em} & Pure shear 2                                    \\ \bottomrule
\end{tabular}
\caption{Decomposition of stress and strain tensors.}
\label{change_basis}
\end{table}

The rank-4 elasticity tensor $C_{ijkl}$ becomes a $4\times 4$ matrix $C_{\alpha \beta}$ relating the two 4-vectors:
\begin{equation}
    \begin{pmatrix}
        \raisebox{-\raisedepth}{\includegraphics[height=.9em]{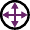}}\\
        \raisebox{-\raisedepth}{\includegraphics[height=.9em]{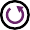}}\\
		\raisebox{-\raisedepth}{\includegraphics[height=.9em]{figures/icons_s2.pdf}}\\
		\raisebox{-\raisedepth}{\includegraphics[height=.9em]{figures/icons_s3.pdf}}
	\end{pmatrix}
    =
    \begin{pmatrix}
      p^{0}\\
      \tau^{0}\\
      s_1^{0}\\
      s_2^{0}\\
    \end{pmatrix}
    +
    \begin{pmatrix}
{\color[rgb]{0.028509,0.250925,0.501969}B} & {\color[rgb]{0.910477,0.739775,0.056336}\Lambda} & {\color[rgb]{0.498037,0.498052,0.498043}C_{02}} & {\color[rgb]{0.498037,0.498052,0.498043}C_{03}} \\
{\color[rgb]{0.501963,0.000037,0.250983}A} & {\color[rgb]{0.065395,0.501825,0.251003}\Gamma} & {\color[rgb]{0.498037,0.498052,0.498043}C_{12}} & {\color[rgb]{0.498037,0.498052,0.498043}C_{13}} \\
{\color[rgb]{0.498037,0.498052,0.498043}C_{20}} & {\color[rgb]{0.498037,0.498052,0.498043}C_{21}} & {\color[rgb]{0.295347,0.736798,0.752687}G_{1}} & {\color[rgb]{0.599997,0.600015,0.600005}K^e+}{\color[rgb]{0.851358,0.364682,0.482766}K^o}   \\
{\color[rgb]{0.498037,0.498052,0.498043}C_{30}} & {\color[rgb]{0.498037,0.498052,0.498043}C_{31}}	 & {\color[rgb]{0.599997,0.600015,0.600005}K^e-} {\color[rgb]{0.851358,0.364682,0.482766}K^o}  & {\color[rgb]{0.295347,0.736798,0.752687}G_{2}}
\end{pmatrix}
	\begin{pmatrix}
		\raisebox{-\raisedepth}{\includegraphics[height=.75em]{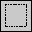}}\\
		\raisebox{-\raisedepth}{\includegraphics[height=.9em]{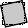}}\\
        \raisebox{-\raisedepth}{\includegraphics[height=.75em]{figures/icons_e2.pdf}}\\
		\raisebox{-\raisedepth}{\includegraphics[height=.9em]{figures/icons_e3.pdf}}
	\end{pmatrix}.
 \label{constitutive_full}
\end{equation}
The constant 4-vector on the right of the equality refers to the pre-stresses present in the material in the absence of any applied deformations, with $p^{0}$, $\tau^{0}$, $s_1^{0}$, and $s_2^{0}$ corresponding pressure, torque, and shear stress 1 and 2 respectively.
The system that we consider is anisotropic due to the driving oscillatory shear in the $x$ direction, so the elasticity tensor can in principle have 16 distinct non-zero components. The entry ${\color[rgb]{0.028509,0.250925,0.501969}B}$ is the bulk modulus, the quantities ${\color[rgb]{0.295347,0.736798,0.752687}G_1}$ and ${\color[rgb]{0.295347,0.736798,0.752687}G_2}$ are the shear modulus, and ${\color[rgb]{0.851358,0.364682,0.482766}K^o}$ is the odd modulus. In the case of isotropy, the elasticity tensor becomes simplified as ${\color[rgb]{0.295347,0.736798,0.752687}G_1} = {\color[rgb]{0.295347,0.736798,0.752687}G_{2}}$, ${\color[rgb]{0.599997,0.600015,0.600005}K^e} = 0$, and all entries labelled ${\color[rgb]{0.498037,0.498052,0.498043}C_{\alpha \beta}}$ vanish \cite{scheibner_odd_2020, fruchart_odd_2023, braverman_topological_2021}.

\subsection{Measurement protocol and results}
\label{protocol}
\begin{figure}[h!]
\centering
\includegraphics[width = .75\columnwidth]{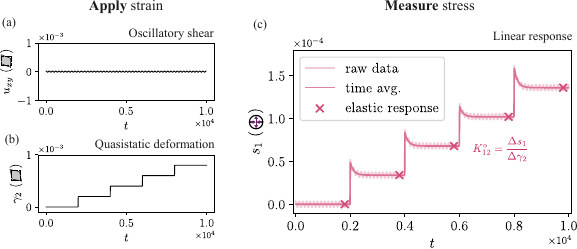}
\caption{Measurement protocol. (a) Time series of the strain amplitude $u_{xy}$ associated with the simple oscillatory shear. (b) Time series of the quasistatic step strains applied to measure the elastic moduli. The magnitude $\gamma_2$ refers to the component of the deformation in the $e_3$ direction. (c) Representative illustration of the typical stress response as a function of time, when applying the quasistatic step strains.}
\label{protocol_schematic}
\end{figure}

In this section, we detail the protocol that we used to measure the elastic moduli shown in Fig. 2 in the main text and show extended results from the moduli measurements. Fig. \ref{protocol_schematic} provides a schematic description of the measurement protocol. 

The system is being driven via oscillatory shear along the $x$ direction, shown in Fig. \ref{protocol_schematic}(a). To probe the linear response of the system, we apply small uniform deformations to the system on top of the oscillatory shear, and measure the stress response. We suppress viscous effects by applying the deformations \textit{quasistatically} \cite{maloney_amorphous_2006} as shown in Fig. \ref{protocol_schematic}(b). Each strain step is accomplished by instantaneously deforming particle positions by an additional magnitude $\Delta \gamma_2$ along the direction of \raisebox{-\raisedepth}{\includegraphics[height=.9em]{figures/icons_e3.pdf}}. After each strain step of $\Delta \gamma_2 \sim 10^{-4}$, the system is allowed to relax for a period of time much longer than each driving cycle before the next strain step is applied. The coarse-grained stress is measured using the Irving-Kirkwood formula \cite{irving_statistical_1950,yang_generalized_2012}
\begin{equation}
    \sigma_{ab} = -\frac{1}{2\mathcal{A}}\sum_{i=1}^N \, \sum_{j\neq i}\,  f_{ij}^a \, r_{ij}^b
    \label{IK}
\end{equation}
where $\mathcal{A}$ is the area of the box, $a,b$ index the Cartesian axes, and $i,j$ are particle indices. A typical stress response of the system, depicted in Fig. \ref{protocol_schematic}(c), shows two features that require additional consideration before proceeding to the calculation of moduli. First, the magnitude of the stress oscillates about a mean value in accord with the applied oscillatory shear. We filter out the oscillating response by first subtracting the stress response of baseline system (oscillatory shear without additional deformations), then performing a time-average of the stress response over several driving periods. We will refer to this time-averaged stress in all subsequent discussion of the stress response, unless otherwise specified. The second feature is the transient response following each strain step, which we address by allowing the system to relax and taking the stress measurement at the end of the relaxation time. The elastic moduli are computed as the slope of the stress response over applied strain. Fig. \ref{protocol_schematic}(c) shows, as an example, the response of shear stress 1 (\raisebox{-\raisedepth}{\includegraphics[height=.9em]{figures/icons_s2.pdf}}) to the applied pure shear 2 (\raisebox{-\raisedepth}{\includegraphics[height=.9em]{figures/icons_e3.pdf}}), corresponding to the modulus $K^o$; other elements of the elasticity tensor in Eq.~(\ref{constitutive_full}) can be measured using the same protocol by taking the appropriate combinations of applied strain and measured stress. \\

\begin{figure}[h!]
\centering
\includegraphics[width = .8\columnwidth]{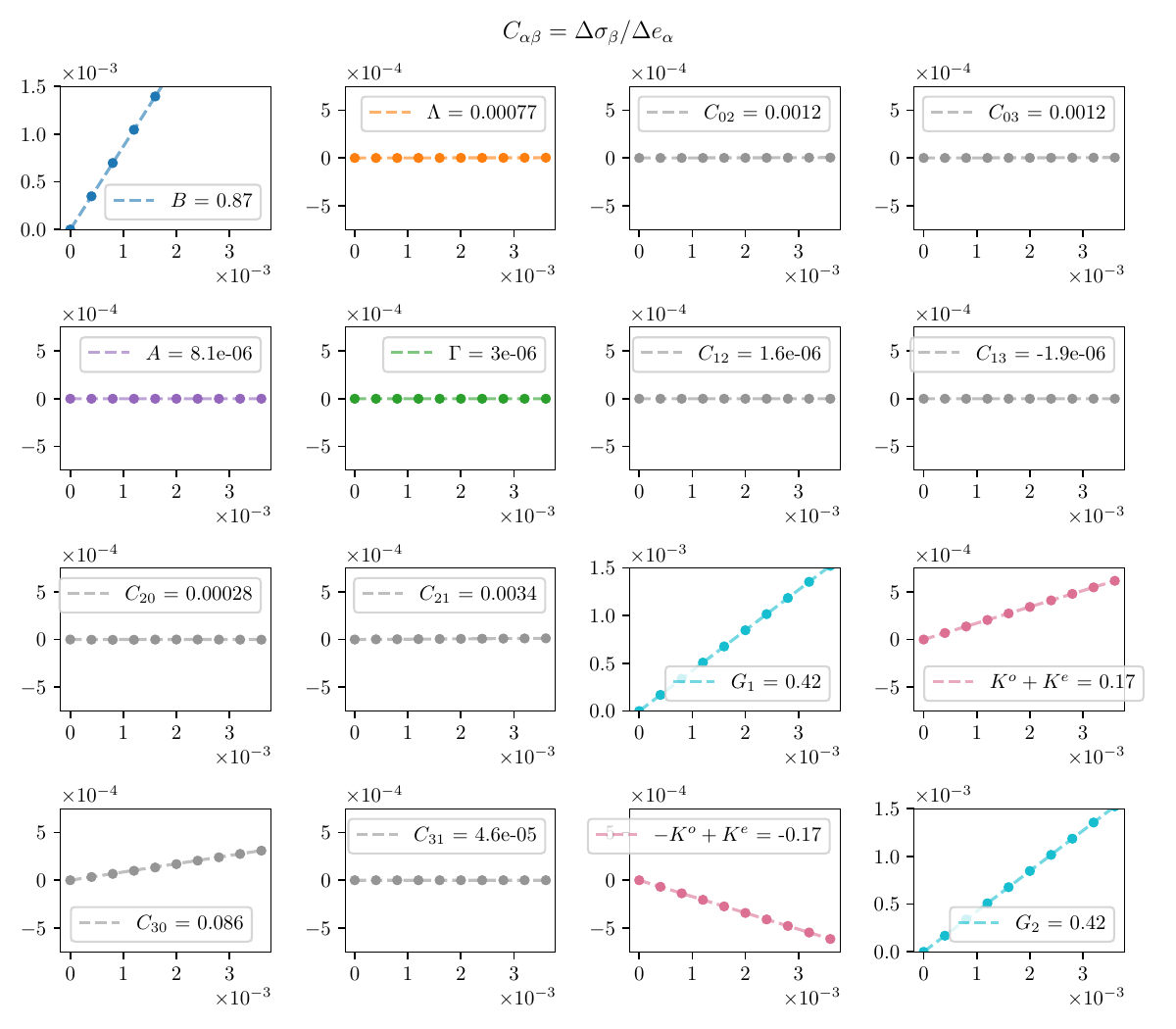}
\caption{Full measurements of $C_{\alpha\beta}$ for the driven $\eps = 0.6$ system with $\gamma_d = 10^{-3}$.}
\label{constitutive_full_data}
\end{figure}
We now present in Fig. \ref{constitutive_full_data} complete moduli measurements for the driven chiral system with $\eps = 0.6$, a part of which is shown in Fig. 2(b) of the main text. This system is in a triangular lattice with periodic boundaries and driven by the SLLOD algorithm. As discussed in the main text, the effects of anisotropy are small in this system due to the relatively weak driving. In the $2\times2$ subspace of pure shears, we see that both pure shear 1 or 2 result in the same even and odd responses such that ${\color[rgb]{0.295347,0.736798,0.752687}G_{1}} = {\color[rgb]{0.295347,0.736798,0.752687}G_{2}}$ and ${\color[rgb]{0.599997,0.600015,0.600005}K^e} = 0$ within measurement error. We note that the only signature of anisotropy in this system's elastic response is found in the small but non-vanishing entry ${\color[rgb]{0.498037,0.498052,0.498043}C_{30}}$ , representing the response of pure shear 2 to isotropic dilation/contraction.\\

When driven more strongly, the system's response deviates from isotropy. The following shows the response of a system that is driven at 10 times greater amplitude compared to the one shown above. We find that in this case ${\color[rgb]{0.295347,0.736798,0.752687}G_1} \neq {\color[rgb]{0.295347,0.736798,0.752687}G_{2}}$, ${\color[rgb]{0.599997,0.600015,0.600005}K^e} \neq 0$, and the entries ${\color[rgb]{0.498037,0.498052,0.498043}C_{30}}$ and ${\color[rgb]{0.498037,0.498052,0.498043}C_{20}}$ acquire magnitudes comparable to the other non-vanishing entries.
\begin{figure}[h!]
\centering
\includegraphics[width = .8\columnwidth]{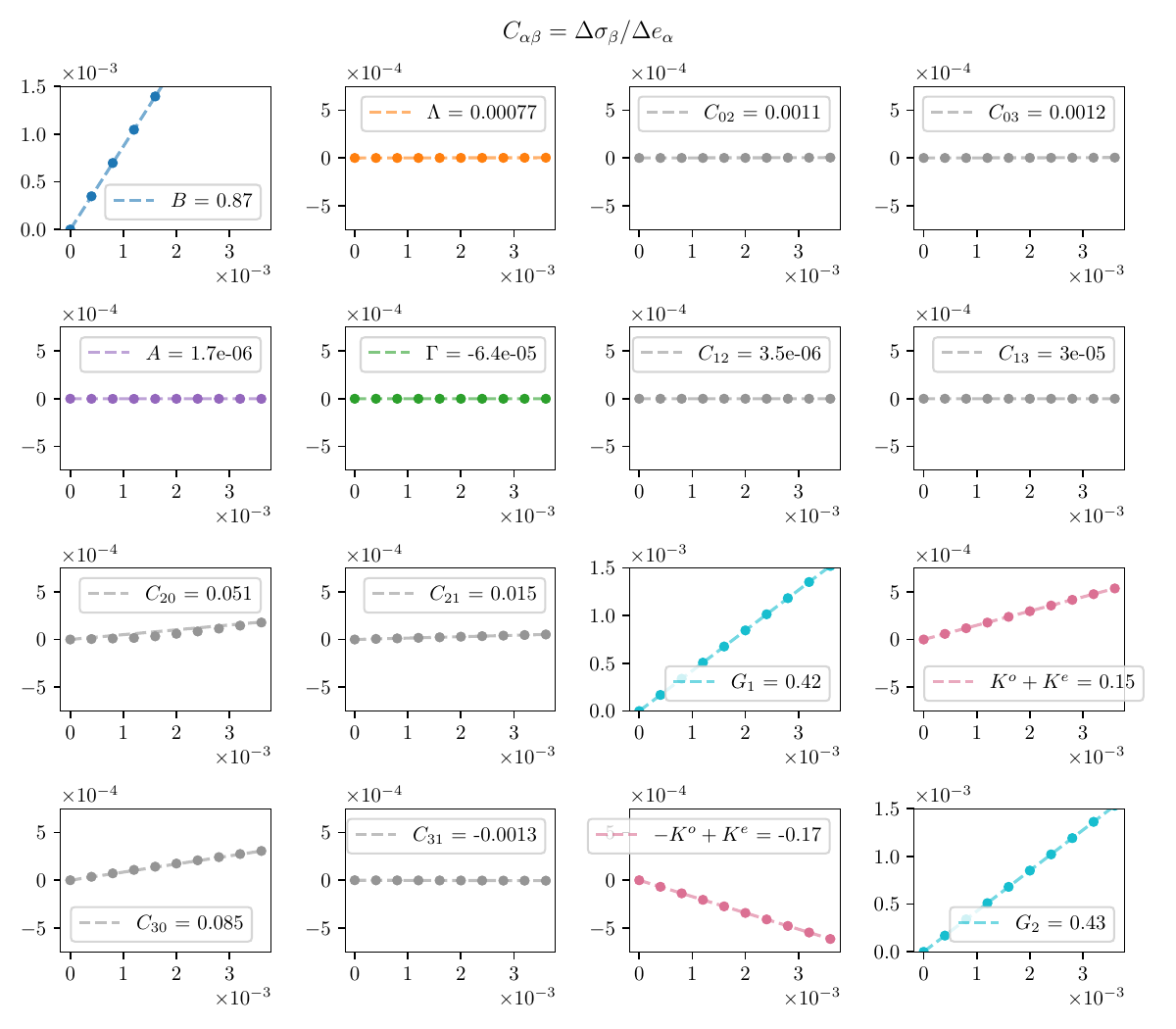}
\caption{Full measurements of $C_{\alpha\beta}$ for the driven $\eps = 0.6$ system with $\gamma_d = 10^{-2}$.}
\label{constitutive_full-large_data}
\end{figure}

\subsection{Effective interactions}
The elastic moduli we measure emerge from the interactions of chiral friction particles under oscillatory shear (described by Eqs. (1)-(3) and Eqs. (\ref{fn_ij})-(\ref{tau_tot})). These equations describe  interactions that are pairwise and instantaneous, and whose transverse part arises from friction and is history dependent (see \ref{1_contact}), precluding a closed form description of the moduli in terms of these forces. This motivates us to ask whether there are time-independent \textit{effective} interaction forces that capture the time-averaged elastic response. These effective interaction forces need not be pairwise even if they emerge from the instantaneous pairwise forces.\\

In the above measurements, we notice that the modulus $A$ coupling isotropic compression to torque density is 0, despite the presence of a nonzero odd shear modulus $K^o$. We conjecture that the continuum level behavior of the system, including its time-averaged elasticity, can be understood from time-independent effective forces. Under this hypothesis: (i) the existence of a nonzero $K^o$ shows that the effective forces must have a transverse component and (ii) $A=0$ shows that the effective forces cannot be fully described as pairwise odd springs, because the Cauchy relations \cite{love_treatise_1892, campanella_note_1994-1} imply that $A=2K^o$ for a triangular lattice with pairwise interactions.\\ 

In order to obtain the elastic moduli from coarse-graining, we make the following additional assumption: the non-pairwise static effective interactions can be seen as strain-dependent pairwise interactions. With this assumption, we can determine an effective pairwise interaction for each imposed strain ($e_0$ to $e_3$ in Table \ref{change_basis}). We can then use the effective interactions to determine the elastic moduli in the corresponding column of the elastic matrix (Eq. (\ref{constitutive_full})). Here, we present microscopic force measurements to substantiate these hypotheses.\\

\begin{figure}[h!]
\centering
\includegraphics[width = .9\columnwidth]{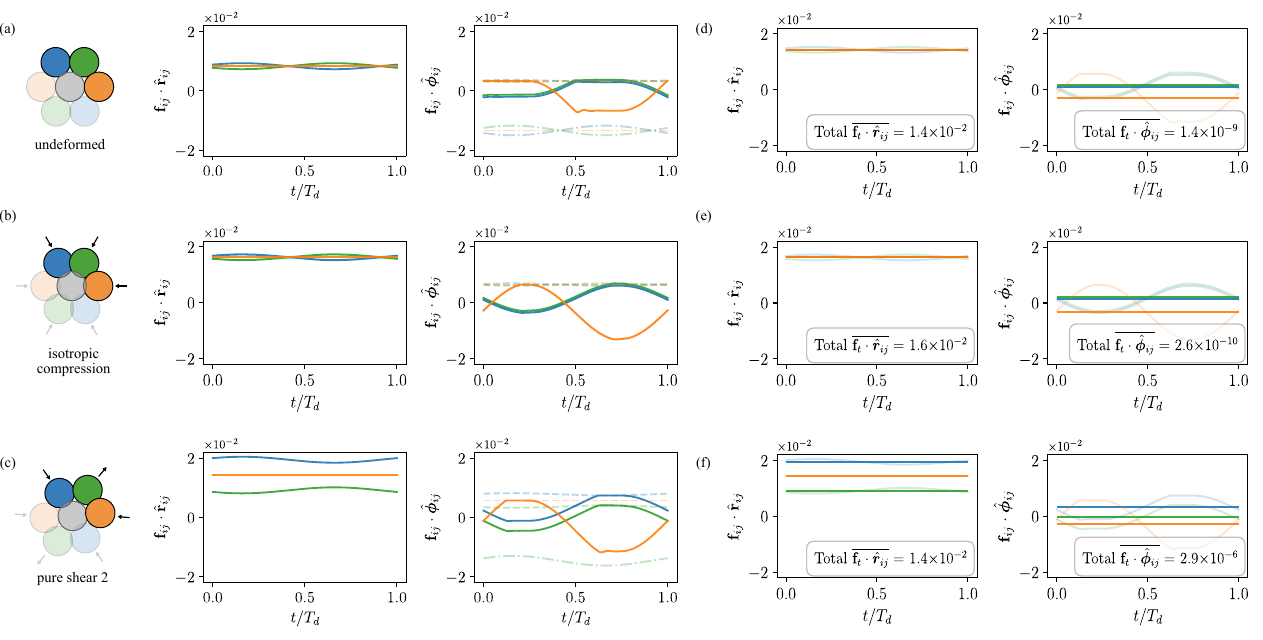}
\caption{Normal and tangential forces $\mathbf{f}_{t_{ij}}$ on the center particle, plotted over a single cycle. The different line colours refer to the different neighbours of the center particle labelled with the same colour (note that by symmetry, there are only three distinct neighbours). In the tangential force plots of (a)-(c), solid lines represent instantaneous tangential forces, dashed lines represent the sliding force condition in the smooth direction $\mu \, (1-\eps) \, |\mb{f}_{n_{ij}}|$, and dash-dotted lines represent the sliding force in the rough direction $\mu \, (1+\eps) \, |\mb{f}_{n_{ij}}|$. In the plots (d)-(f), opaque solid lines represent cycle averages while transparent solid lines represent instantaneous forces. Measurements performed under (a), (d) no deformation (b), (e) isotropic compression $e_0$ and (c), (f) pure shear $e_3$. }
\label{2-ft_v_strain}
\end{figure}

Fig. \ref{2-ft_v_strain} shows plots of the normal and tangential forces on a particle in the triangular lattice over each driving cycle. Three strain configurations are shown: undeformed, under isotropic compression $e_0$, and under pure shear $e_3$. In the undeformed configuration in Fig. \ref{2-ft_v_strain}(a) and (d), the normal force magnitude oscillates over a cycle around a mean of $1.4 \times 10^{-2}$. During parts of the cycle, we see in Fig. \ref{2-ft_v_strain}(a) that the center particle's neighbours are \textit{sliding} against the center particle in the smooth direction, appearing as constant sections in the tangential force plots where $|\mb{f}_{t_{ij}}| = \mu \, (1-\eps) \, |\mb{f}_{n_{ij}}|$. The cycle-averaged tangential force on the center particle summed over all neighbours is, however, negligible (Fig. \ref{2-ft_v_strain}(d), right). In light of Eq. (\ref{IK}), and noting that here the substrate resistance is set to $\zeta, \zeta^r = 0$, this corresponds to the condition that the system's time-averaged torque density vanishes in the undeformed state. (See the following subsection for additional discussion.) \\ 

Next, we can consider the forces on the center particle over each cycle under strained configurations. To understand the odd moduli $A$ and $K^o$, we focus on isotropic compression and pure shear of type 2 respectively; similar analysis can be performed for all four deformations in Table \ref{change_basis}. When isotropic compression is applied, Fig. \ref{2-ft_v_strain}(b) and (e) show that the normal forces on the center particle increase on average. The sliding friction forces increase in proportion to the normal forces via $\mu \, (1-\eps) \, |\mb{f}_{n_{ij}}|$, but the average values of the tangential forces are unchanged. The time-averaged tangential forces (and therefore torque density) remain vanishing resulting in $A$ =0.\\

Upon application of pure shear 2 (Fig. \ref{2-ft_v_strain}(c) and (f)), the change in the normal forces due to each neighbor depends on whether it lies along the compression or the expansion axis, relative to the center particle. This corresponds via Eq.~(\ref{IK}) to an increase in shear stress $\sigma_3$, leading to a finite shear modulus $G$. The sliding friction forces $\mu \, (1-\eps) \, |\mb{f}_{n_{ij}}|$ change in proportion to the normal forces. In this case, however, the cycle-averaged tangential force due to each neighbour shifts along with the sliding forces. Applying Eq. (\ref{IK}) to the cycle-averaged tangential forces after the shift results in an increase in shear stress $\sigma_2$ with magnitude $\mu \, (1-\eps)$ times the shear stress $\sigma_3$. This finally leads to an odd shear modulus $K^o\approx \mu \, (1-\eps)\, G$.\\

We can summarize the measurement results as follows. The undeformed system behaves as if each grain is under an effective microscopic force that features a normal force but no effective tangential force. Under each of the four distinct orthogonal deformations $e_\alpha$ (see Table \ref{change_basis}), the system responds as if the grains are under potentially distinct effective tangential forces $f_t^{\textrm{eff}}(r)$. For example, our measurements show that
\begin{align}
    f_t^{\textrm{eff}}(r) \approx \begin{cases}
        0 & \text{under isotropic compression $e_0$}\\
        \mu\,(1-\eps)\big(\Delta \overline{\abs{\mathbf{f}_n(r)}}\big) & \text{under pure shear $e_3$}
    \end{cases},
    \label{ft_eff}
\end{align}
where $\Delta \overline{\abs{\mathbf{f}_n(r)}} = \big(\overline{\abs{\mathbf{f}_n(r)}} - \overline{\abs{\mathbf{f}_n(a)}}\big)$ refers to the change in the average normal forces from the undeformed configuration. Analogous estimates for deformations $e_1$ and $e_2$ can be obtained through similar analyses of microscopic forces. By coarse-graining \cite{braverman_topological_2021, fruchart_odd_2023} the effective tangential forces in Eq. (\ref{ft_eff}), we obtain odd moduli that correspond to each of the applied deformations. Namely, the modulus $A$ = 0 follows from coarse-graining the corresponding $f_t^{\textrm{eff}}(r) \approx 0$, and the modulus $K^o\approx \mu \, (1-\eps)\, G$ follows from $f_t^{\textrm{eff}}(r) \approx\mu\,(1-\eps)\big(\Delta \overline{\abs{\mathbf{f}_n(r)}}\big) $.\\

Finally, we remark that Eq.~(\ref{ft_eff}) demonstrates that a pairwise description of effective microscopic forces is insufficient. The presence of different effective tangential forces \textit{dependent on the deformation applied} corroborates the inference from Cauchy relations that a non-pairwise description is necessary to fully capture the behaviour of this system.

\subsection{Torque density}
This section investigates in the observation of vanishing $A$ and torque density. From the Irving-Kirkwood formula (Eq.~(\ref{IK})), the torque density in the system is given by
\begin{equation}
    \sigma_1 \equiv \frac{1}{2} (\sigma_{yx} -\sigma_{xy}) =  \frac{1}{4\mathcal{A}}\sum_{i=1}^N \, \sum_{j\neq i}\,  \left( \mb{f}_{ij} \times \mb{r}_{ij} \right) \cdot \hat{\mb{z}}.
    \label{sigma_1}
\end{equation}
A comparison between Eq. (\ref{sigma_1}) and Eq. (\ref{tau_tot}), along with the fact that the torque on each particle is the same in the ideal triangular lattice with periodic boundary conditions, shows that the torque density is proportional to the interaction torque on each particle $\sigma_1  = -N\tau_i^{\text{int}}/({2\mathcal{A}})$ (note that this section uses a sign convention that clockwise angular velocity is positive). On the other hand, the equation of motion for the angular degrees of freedom is provided in Eq. (2) of the main text, which we can write as
\begin{equation}
    I_i \, \dee_t \omega_i = \tau_i^{\text{int}} + \tau_i^{\text{ext}}
    \label{Omega_eom}
\end{equation}
where $\omega_i(t)$ is the instantaneous angular velocity of particle $i$, $\tau_i^{\text{int}}$ is the total interaction force due to other particles, and $\tau_i^{\text{ext}}$ refers to all other torques including substrate resistance $-\zeta^r \dot{\theta_i}$ and possibly external torques directly applied to the particles. In all the measurements presented in Figs. \ref{constitutive_full_data}-\ref{2-ft_v_strain}, the substrate resistance $-\zeta^r$ is set to 0, and we apply no direct external torques. Under these conditions, the observation that the particles reach a steady state with a well-defined averaged spinning speed $\Omega_i = \overline{\omega_i}$ indicates that the time-averaged torque must vanish:
\begin{equation}
    0 = I_i  \, \dee_t\Omega_i  = I_i  \, \overline{\dee_t \omega_i} = \overline{\tau_i^{\text{int}}}.
    \label{Omega_eom_ss}
\end{equation}
The frictional interactions, which are responsible for driving the particles to spin and also for resisting their spinning, balance out over each cycle in the steady state. The time-averaged torque density, computed from the Eq. (\ref{sigma_1}) as $\overline{\sigma_1}  = -N\overline{\tau_i^{\text{int}}}/({2\mathcal{A}})$, therefore vanishes in the steady state.\\

\begin{figure}[h!]
\centering
\includegraphics[width = .6\columnwidth]{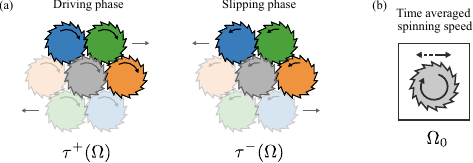}
\caption{ (a) Schematic of the first and the second halves of a driving cycle. Color labels indicate the three distinct neighbours of the center particle, and gray arrows represent the translation motion during the phases of the driving cycle. (b) The time-averaged spinning speed $\Omega_0$ at steady state. }
\label{2-torque}
\end{figure}

Moreover, the transduction mechanism from driving oscillatory shear to particle rotation ensures that a steady state is always eventually reached, even in the absence of substrate resistance. To rationalize this, let us define $\tau^+(\Omega)$ to be the averaged total interaction torque on each particle during the first ``driving" half of the cycle, and likewise define $\tau^-(\Omega)$ for the second ``slipping" half of the cycle. This is shown in Fig. \ref{2-torque}, where the three distinct neighbours of the center particle (labelled by colors blue, green, and orange) are responsible for the driving and slipping torques on the center particle. The sign convention for $\tau^\pm$ is chosen so that $\tau^+>0$ leads to $\dee_t\Omega>0$ while $\tau^->0$ leads to $\dee_t\Omega<0$, so the time-averaged torque over a full driving cycle is $\overline{\tau_i^{\text{int}}} = (\tau^+ - \tau^-)/2$. \\

The driving torque $\tau^+(\Omega)$ generally decreases with the average spinning speed $\Omega$, while $\tau^-(\Omega)$ tends to increase with $\Omega$, and it is this feedback mechanism that produces a steady state in $\Omega$. The interactions between neighbours in the same layer (labelled by orange in Fig. \ref{2-torque}) always counteract the spinning direction $\Omega$, so it suffices to consider interactions between neighbours in a different layer (labelled by blue and green). For instance, suppose the system begins at rest with $\Omega = 0$. The direction of relative motion between the contact points indicates that neighbours in a different layer can exert a larger (smaller) frictional force during the driving (slipping) phase. The imbalance $\tau^+>\tau^-$ increases the average spinning speed $\Omega$. However, as $\Omega$ increases, the relative speed of the contact points decreases (increases) during the driving (slipping) phase. For large enough $\Omega$, the relative motion in the driving phase will eventually reverse, and $\tau^+$ changes from positive to negative sign. The slipping torque $\tau^-$ experiences no such sign change, as it increases with $\Omega$ until it saturates when $\omega_i(t)$ becomes always counterclockwise. The total torque $\overline{\tau_i^{\text{int}}}(\Omega)$ therefore crosses $0$ with a decreasing slope at some value $\Omega_0$. The relation near this value $\overline{\tau_i^{\text{int}}} = I_i  \, \dee_t\Omega_i \approx \alpha (\Omega-\Omega_0)$ where $\alpha = d(\overline{\tau_i^{\text{int}}})/d\Omega\bigr|_{\Omega_0} < 0$ indicates that the system tends towards the stable fixed point $\Omega = \Omega_0$.

The above analysis addresses the vanishing of torque density at steady state in the system when Eq. (\ref{Omega_eom}) involves only interaction torques. Finally, we present an example to illustrate the effect of $\tau_i^{\text{ext}}  =  -\zeta^r \dot{\theta_i}$.
\begin{figure}[h!]
\centering
\includegraphics[width = .8\columnwidth]{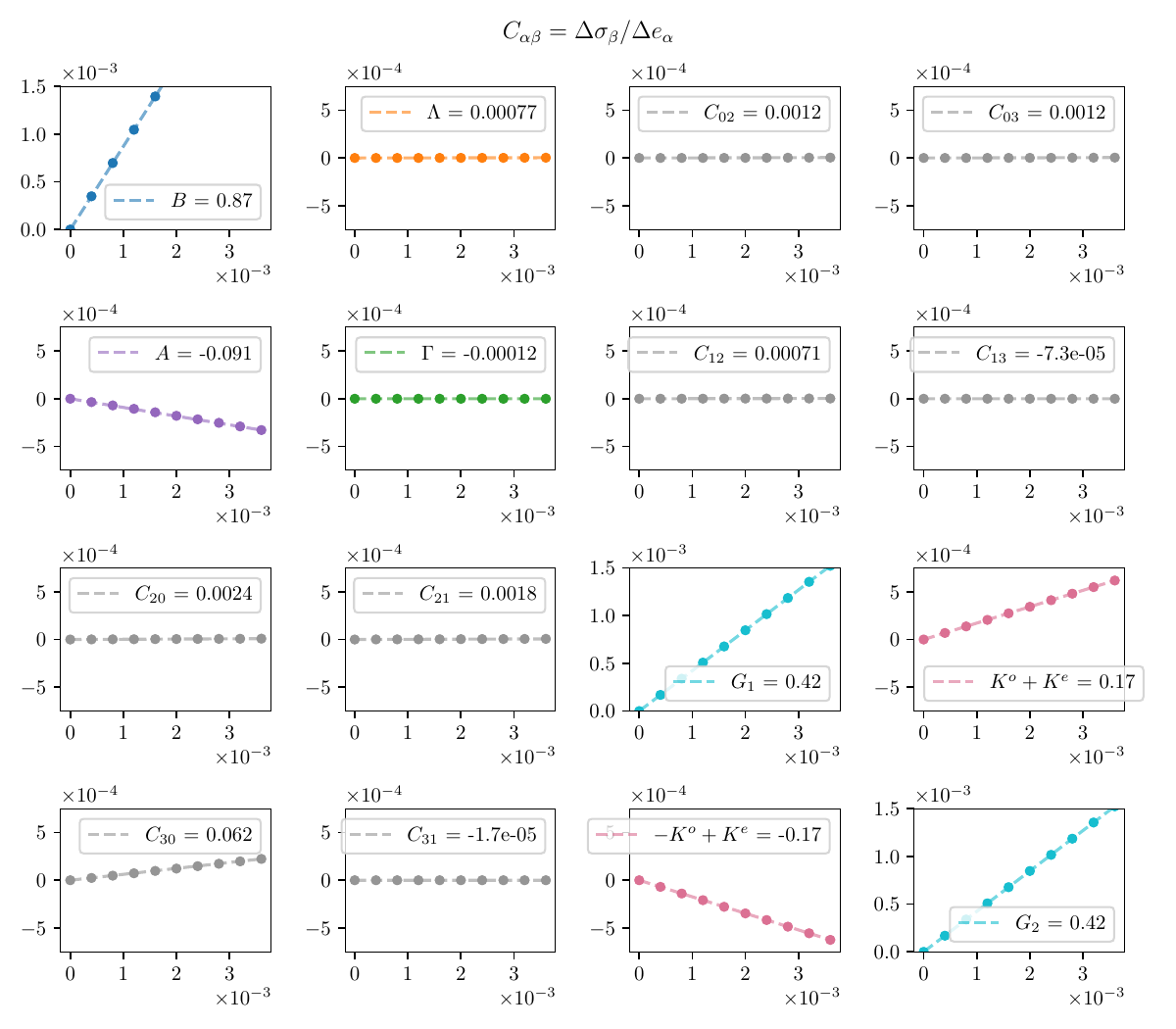}
\caption{Full measurements of $C_{\alpha\beta}$ for the driven $\eps = 0.6$ system with $\gamma_d = 10^{-3}$, with substrate resistance $\zeta, \zeta^r = 100$.}
\label{constitutive_full_zf}
\end{figure}
In Fig. \ref{constitutive_full_zf}, we perform the same measurements of the moduli as shown in Fig. \ref{constitutive_full_data}, now including a substrate resistance of $\zeta^r = 100$ in the system. The odd shear modulus $K^o$ is unaffected, but the modulus $A$ is now finite, indicating the presence of a torque density that changes upon dilation. The system is still being observed at a steady state value of $\Omega_i$, but in this case the right hand side of Eq. (\ref{Omega_eom_ss}) includes an extra term $\overline{\tau^\text{ext}} = -\zeta^r \Omega_i$. Since the torque density calculated from the Irving-Kirkwood formula in Eq. (\ref{sigma_1}) accounts only for the interaction torques, the presence of the $\overline{\tau^\text{ext}}$ now leads to a non-vanishing (negative) value of torque density $\sigma_1$ at the steady state, which we measure to be $\sigma_1 = -4.8\times 10^{-4}$. This corresponds to the constant pre-stress term in Eq. (\ref{constitutive_full}). The same reasoning applies when there are other forms of external torques, such as particles directly driven to spin by external fields or internal sources of energy; it is the choice to ignore such external torques when computing the torque density by Eq. (\ref{sigma_1}) that results in a non-vanishing value of $\sigma_1$ at the steady state. In our system, the measurement of the modulus $A$ in Fig. \ref{constitutive_full_zf} plots the change in $\sigma_1$ as the system is isotropically dilated, showing that the magnitude of the torque density increases upon dilation as indicated by $A<0$. This is consistent with the dependence of average spinning speed $\Omega$ on the area fraction of the system, shown in main text Fig. 3(b). The increase in $\Omega$ with dilation leads to an increasing magnitude of $\overline{\tau^\text{ext}}$ due to substrate resistance, which at the steady state implies an increasingly negative value of the torque density $\sigma_1$.

\subsection{Consequences of odd elasticity}
Many analytical results follow immediately from the presence of the odd modulus $K^o$ in the elasticity tensor. We examine two such consequences: asymmetry of the elastic Green's function and shear cycles that extract work.
\subsubsection{Green's function for linear elasticity}
The elastic Green's function $G_{ij}(\mb{r})$ is defined as the $i^{\text{th}}$ component of the displacement field caused by a unit point force in the $j^{\text{th}}$ direction. Here, we will consider the 2D isotropic Green's function for the point force $\mathbf{F} \delta^2(\mathbf{r})$ applied at the origin of an infinite medium, and solve for the resulting strain field $u_i$. The equilibrium equation for the continuum is $\dee_j\sigma_{ij} = \dee_j C_{ijkl} u_{kl} = -F_j$, for which we use the tensor expression for $C_{ijkl}$ found in Refs. \cite{scheibner_odd_2020, braverman_topological_2021, fruchart_odd_2023} and write (sums over repeated indices implied)
\begin{equation}
    \left(B \dee_i \dee_j  + (G\delta_{ij} + K^o \epsilon_{ij}) \dee_m\dee_m\right) u_j = -F_i \delta^2(\mathbf{r}).
    \label{green_de}
\end{equation}

The modulus $A$ has been omitted from this calculation, since it is typically measured to be vanishing in our system. To solve this equation for the displacements $u_j$, perform a Fourier transform to obtain
\begin{equation}
    \left(B k_i k_j  + (G\delta_{ij} + K^o \epsilon_{ij}) k^2\right) \Tilde{u}_j = F_i,
\end{equation}
which gives the following expression for $\Tilde{u}_j$ when inverted:
\begin{equation}
    \Tilde{u}_i(\mb{k}) = \frac{1}{M^2k^2} \left[ G \delta_{ij} - K^o \epsilon_{ij} + \frac{B}{(M^2+G B) k^2} (- G^2 k_ik_j + G K^o (\epsilon_{il}k_l k_j - \epsilon_{jl}k_l k_i) + (K^o)^2 \epsilon_{il}\epsilon_{jm}k_l k_m )\right] F_j
\end{equation}
where we have defined $M^2 = G^2 + (K^o)^2$. The inverse Fourier transform of this expression leads to the solution for the displacements:
\begin{align}
    u_i(\mb{r}) = \frac{F_j}{2 \pi M}\Bigg[&\big(G \delta_{ij} - K^o \epsilon_{ij}\big) \log\big(\tfrac{r}{a}\big) \nonumber\\
    &+\left(\tfrac{B}{4M+ 4G B}\right) \Big(-G^2 \delta_{ij} + 2G K^o \epsilon_{ij} + (K^o)^2 \delta_{ij}\Big) \Big(2\log(\tfrac{r}{a}) -1\Big) \nonumber\\
    &+ \left(\tfrac{B}{4M+ 4G B}\right) \Big(-G^2 r_i r_j \delta_{ij} + G K^o (\epsilon_{il}r_l r_j - \epsilon_{jl}r_l r_i) + (K^o)^2 \epsilon_{il}\epsilon_{jm}r_l r_m\Big) \Big(\tfrac{2}{r^2}\Big) \Bigg].
\end{align}
We observe that the presence of $K^o$ causes the displacement field to rotate relative to the direction of the point force. Explicitly, for a point force in the $x$ direction at the origin  we obtain the following components of the displacement field:
\begin{align}
     u_x(\mb{r}) = \frac{F_j}{2 \pi M}\Big[&G \log\big(\tfrac{r}{a}\big) +\left(\tfrac{B}{4M+ 4G B}\right) \Big(-G^2 + (K^o)^2 \Big) \Big(2\log(\tfrac{r}{a}) -1\Big) + \left(\tfrac{B}{4M+ 4G B}\right) \Big(-G^2 x^2 + (K^o)^2 y^2 \Big) \Big(\tfrac{2}{r^2}\Big) \Big]\nonumber\\
     u_y(\mb{r}) = \frac{F_j}{2 \pi M}\Big[&K^o \log\big(\tfrac{r}{a}\big) +\left(\tfrac{B}{4M+ 4G B}\right) \Big(-2G K^o\Big) \Big(2\log(\tfrac{r}{a}) -1\Big) + \left(\tfrac{B}{4M+ 4G B}\right) \Big(-G^2 + 2G K^o + (K^o)^2 \Big) \Big(\tfrac{2xy}{r^2}\Big) \Big].\label{green_u}
\end{align}

The $\log(r/a)$ dependence (where $a$ is a cutoff distance on the order of a lattice spacing) originates from the 2D Green's function for the Laplace operator $\dee_m\dee_m$. In practice, the assumption of infinite medium boundary conditions does not match our numerical experiments, which feature  periodic boundary conditions. Thus, rather than directly comparing Eq. (\ref{green_u}) to the simulation results in main text Fig. 1(c), we instead extract from Eq. (\ref{green_u}) the average angle by which the displacement field rotates relative to the direction of the point force:
\begin{equation}
    \langle u_\phi \rangle \equiv \frac{1}{L_x L_y}  \int_{-L_x/2}^{L_x/2} \int_{-L_y/2}^{L_y/2}  \arctan\left(\frac{u_y(\mb{r})}{u_x(\mb{r})}\right) \, dx\,dy,.
\end{equation}
This is a quantity which depends on the moduli $B$, $G$, and $K^o$. By performing the calculation using the moduli values measured for our system (Figs. 2 and \ref{constitutive_full_data}), we can compare the prediction Eq. (\ref{green_u}) to the behaviour of the strain fields observed in our numerical experiment in Fig. 1(c)  of the main text. This comparison is shown in Fig. 2(c).

\subsubsection{Elastic engine cycles}
\begin{figure}[]
\centering
\includegraphics[width = .25\columnwidth]{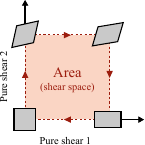}
\caption{Cycle of pure shears.}
\label{ucyc}
\end{figure}
In Fig. 2(d), we perform a quasistatic shear cycle as depicted schematically in Fig. \ref{ucyc}, and compute the work extracted from the system in the process. The work per unit area done by a solid due to a deformation cycle is
\begin{align}
    dW &= \sigma_{ij} \,d u_{ij} = \tfrac{1}{2} \sigma_{\alpha} \,d e_\alpha\\
    W &= \frac{1}{2}\oint \sigma_{\alpha} \,d e_\alpha
\end{align}
where the $\sigma_{\alpha}$ and $e_\alpha$ are the combinations of stresses and strains given in Table \ref{change_basis}. (For our system, recall that the stresses refer to the time-averaged stresses as described in Section \ref{protocol}.) The odd modulus $K^o$ contributes to the anti-symmetric part of the elasticity tensor (Eq. (\ref{constitutive_full})), which implies the existence of quasistatic strain cycles where the work extracted from the solid is nonzero. We provide a derivation of this result, and refer the reader to Refs. \cite{scheibner_odd_2020, fruchart_odd_2023} for in-depth discussions. Considering the strain cycle in Fig. \ref{ucyc}, we can relate the total work extracted from the system over the cycle the the elastic moduli as follows:
\begin{align}
    W &= \frac{1}{2}\oint \sigma_{\alpha} \,d e_\alpha \\
    &= \frac{1}{2}\oint C_{\alpha\beta}e_\beta \,d e_\alpha \\
    &= \frac{1}{2}\oint \big(G\delta_{\alpha\beta}e_\beta + K^o \epsilon_{\alpha\beta}e_\beta\big) \,d e_\alpha\\
    &= \frac{1}{2}\iint \epsilon_{\alpha\gamma} \dee_\gamma \big(G\delta_{\alpha\beta}e_\beta + K^o \epsilon_{\alpha\beta}e_\beta\big) \,d^2 e \quad\quad\text{ Stokes' Theorem}\\
    &= K^o \times (\text{Area in shear space}).
\end{align}

The work gained from performing the quasistatic shear cycle is nonzero and proportional to the odd modulus $K^o$. Note that the sign of the work depends on the direction in which the shear cycle is carried out, with the sign reversing when the shear cycle is performed with opposite handedness. In Fig. 2(d) of the main text, we measure the work extracted from strain cycles of varying areas, and compare the resulting dependence to the relation derived above.

\subsection{Driving from the boundary}
\begin{figure}[h!]
\centering
\includegraphics[width = .8\columnwidth]{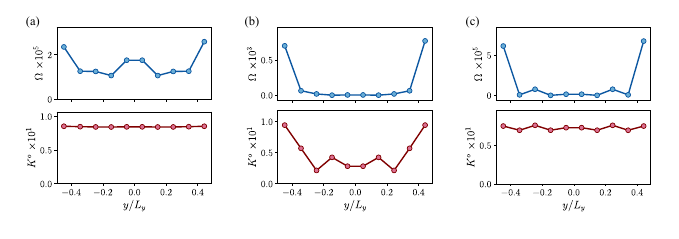}
\caption{Spatial dependence of $\Omega$ (top) and $K^o$ (bottom) in the system driven from the boundary. The horizontal axis is the $y$ coordinate of each coarse-graining subsystem, divided by the system size $L_y$ in the $y$ direction. The systems are driven with (a) $T_d = 100$, $\gamma_d = 10^{-3}$, (b) $T_d = 20$, $\gamma_d = 10^{-3}$, (c) $T_d = 20$, $\gamma_d = 2\times10^{-4}$.}
\label{2-wdrv}
\end{figure}

In this section, we further investigate the position dependence of the spinning speed and odd modulus in the boundary-driven system. Fig. \ref{2-wdrv} shows $\Omega$ and $K^o$ for a boundary-driven system consisting of 100 layers of particles along the shear gradient direction. The $\Omega$ and $K^o$ values are coarse-grained over 10 layers; in other words, the $\Omega$ data are averaged (emergent) spinning speeds over the 10 layers, and $K^o$ is computed by applying the Irving-Kirkwood formula (Eq. (\ref{IK})) to the 10-layer subsystem.\\

We present data from systems under three driving conditions in Fig. \ref{2-wdrv}. When the driving cycle period is $T_d = 100$ (identical to main text Fig. 3(f)), we find that $\Omega$ can vary significantly across the subsystems, while $K^o$ remains relatively constant. However, when the driving cycle's frequency is 5 times higher, the coarse-grained $K^o$ also shows a wide variation over the different subsystems. If the driving amplitude is also reduced by 5 times, then a greater uniformity in the odd modulus is attained.\\

These data indicate that the spatial dependence of the odd modulus is sensitive to the details of the driving cycle. Typically, we find that the odd modulus becomes non-uniform when the driving cycle exceeds the response time of the system, which can be addressed by a suitable choice of driving cycle parameters relative to the system parameters.

\vfill
\pagebreak
\section{Polycrystalline systems}

In the main text, we investigate the behaviour of the driven chiral solid when grain boundaries are present. We present here a more detailed description of the system, including the emergent spinning speeds and odd moduli, mechanisms causing grain boundary self-healing, and the stability of the solid.

\subsection{Emergent $\Omega$ and odd moduli}
\label{3-measurements}
\begin{figure}[h!]
\centering
\includegraphics[width = .8\columnwidth]{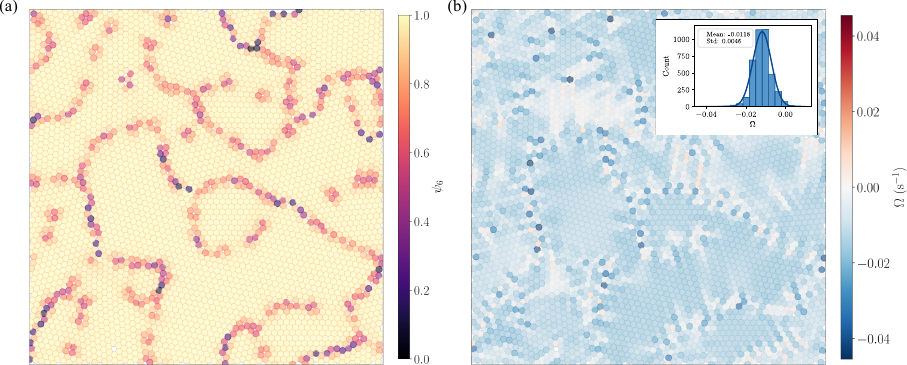}
\caption{Polycrystalline system. Maps of (a) $\psi_6$ and (b) emergent spinning speed $\Omega$. Inset of (b) shows the distribution of spinning speeds.}
\label{3-defect_spin}
\end{figure}

Unlike in the uniform triangular lattice, the inhomogeneous density in the polycrystalline system creates a spread in the emergent spinning speeds $\Omega$ across the different crystalline regions. This is visible in the map of $\Omega$ in Fig. \ref{3-defect_spin}(b). By comparing with the $\psi_6$ plot of Fig. \ref{3-defect_spin}(a), it shows a relatively uniform $\Omega$ within crystalline patches and a wider spread near grain boundaries. The distribution shown in the inset confirms that spinning speeds are distributed around an average that is clockwise, which is consistent with the spinning direction in the triangular lattice for the same $\eps$.\\

\begin{figure}[h!]
\centering
\includegraphics[width = .6\columnwidth]{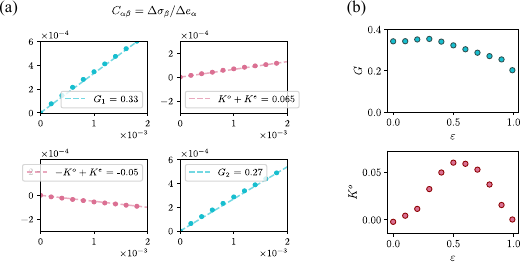}
\caption{Elastic response of the polycrystalline system. (a) Measurements of the shear modulus and odd modulus (for chirality $\eps = 0.6$ and driving frequency $T_d = 1$). (b) Dependence of the shear and odd modulus on the degree of chirality $\eps$. }
\label{3-defect_C}
\end{figure}

We can measure the elastic response of this polycrystalline solid using the same protocol described in Section \ref{protocol}, with a modification of choosing shorter relaxation times after each strain step. Here we use typically around $10$ to $10^2$ times the driving cycle period $T_d$, compared to the $10^3\times T_d$ used in earlier sections. Since the system undergoes slow deformations on its own (main text Fig. 4(a)), this modification allows us to probe the response to the applied deformations rather than its own spontaneous deformations. Over these intermediate time scales, we observe a linear response that is shown in Fig. \ref{3-defect_C}(a). The odd modulus $K^o$ is nonzero in this case; the solid retains its odd elastic properties even in the presence of grain boundaries. We also examine the dependence of the odd modulus $K^o$ on the degree of chirality $\eps$ in the polycrystalline system. As shown in Fig. \ref{3-defect_C}(b), $K^o$ is non-monotonic in $\eps$. This behaviour shows a similar trend to the predictions and observations for the triangular lattice in Fig. 3(c), where $K^o$ is maximized at intermediate values of $\eps$ due to the presence of the strongest cycle-averaged tangential forces.

\subsection{Microscopic mechanisms of defect motion}
In this section, we examine the mechanisms of defect motility that result in the anisotropic grain boundary motion shown in Fig. 3(b)-(c) of the main text. We note that the substrate drag for the simulations in these figures was set to $\zeta, \zeta^r = 1$. Previous work \cite{braverman_topological_2021} on odd elastic solids with defects distinguishes between two classes of forces acting on dislocations. On the one hand, there is the Peach-Koehler force $f^{\text{PK}}_k = \epsilon_{ki} \sigma_{ij} b_j$, which is an estimate of the effective force on a dislocation due to coarse-grained stresses at the continuum level ($\sigma_{ij}$ is the pre-existing coarse-grained stress in the solid, and $b_j$ is the dislocation's Burgers vector). The Peach-Koehler expression is not unique to odd elastic solids. It is a general formula involving only on the stress field and the Burgers vector, and it is independent of how the stress field arises \cite{scheibner_odd_2020, peach_forces_1950, lubarda_dislocation_2019}. On the other hand, continuum considerations become inadequate to describe defect motion when microscopic forces are nonconservative. For such systems, Ref. \cite{braverman_topological_2021} identifies a core force $f^{\text{core}}$ that arises from microscopic work cycles and is unaccounted for in the continuum description of stress. 
The two types of forces can be understood heuristically as follows: $f^{\text{PK}}$ is the force due to bonds not crossing the glide plane, while $f^{\text{core}}$ arises from those bonds that do cross the glide plane \cite{fruchart_odd_2023}. 
We will here demonstrate that depending on the orientation of the glide plane, either of the two situations can dominate in our system. 
\\

\begin{figure}[h!]
\centering
\hspace{-1em}\includegraphics[width = .8\columnwidth]{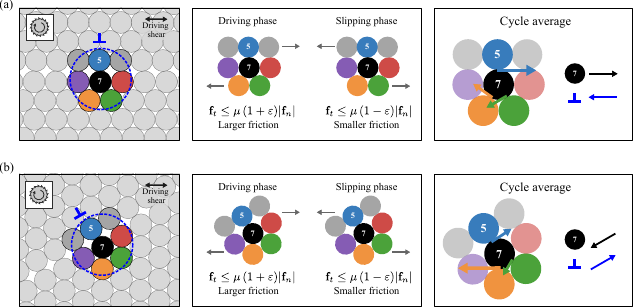}
\caption{Free-body diagram of the interaction of the sevenfold coordinated particle (in black) with neighbours, averaged over a cycle, for Burgers vectors making an angle of (a) $\phi_b = 0^{\circ}$ and (b) $\phi_b = 30^{\circ}$ relative to the shear axis.}
\label{defect_fbd}
\end{figure}

We consider a single dislocation and consider the forces acting on the particles that are part of a dislocation over a driving cycle. An imbalance of these forces would indicate the direction of motion of the defect. Fig. \ref{defect_fbd} demonstrates how such an imbalance arises from the driving oscillatory shear and chiral friction. Focusing on the sevenfold coordinated particle of the defect, we see that the effect of bonds perpendicular to the oscillatory shear axis is enhanced. The frictional forces on the sevenfold particle are asymmetric during the two halves of the driving cycle due to chiral friction. However, the neighbour whose bond is perpendicular to the shear axis experience greater relative motion during the driving cycle, which increases their frictional interactions overall. In addition, the direction of the force is aligned with the glide plane of the defect, so the force is more effective at doing work on the defect. The net force on the sevenfold particle is therefore dominated by the interaction with the neighbour whose bond is perpendicular to the shear axis.\\

\begin{figure}[h!]
\centering
\hspace{-1em}\includegraphics[width = .6\columnwidth]{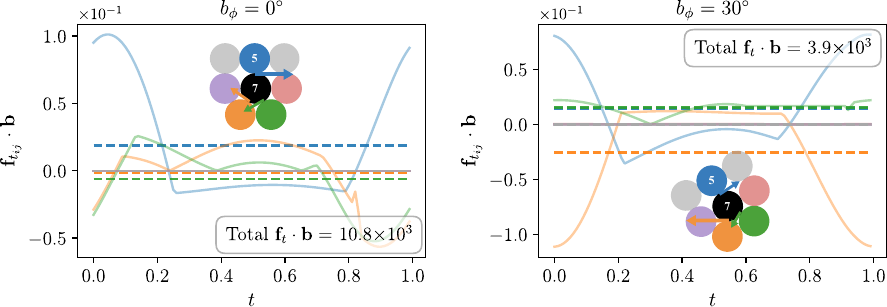}
\caption{Tangential forces $\mathbf{f}_{t_{ij}}$ on the sevenfold particle projected onto the Burgers vector $\mathbf{b}$, plotted over a single cycle. The different line colours refer to the neighbours of the sevenfold particle labelled with the same colour. Solid lines are instantaneous values of $\mathbf{f}_{t_{ij}}\cdot \mathbf{b}$, and dashed lines represent the average value over the cycle. The total $\mathbf{f}_{t}\cdot \mathbf{b}$ is the sum of the cycle-averaged values over all neighbours.}
\label{defect_ft}
\end{figure}

In the configuration shown in main text Fig. \ref{defect_fbd}(a), the interaction of the particle labelled in blue dominates and results in a net force in the $+x$ direction, which corresponds to defect motion in the $-x$ direction. It is straightforward to see how the net force direction can reverse when the defect has a different orientation. Fig. \ref{defect_fbd}(b) shows the case where the Burgers vector is rotated by $\phi_b = 30^{\circ}$ relative to the shear axis. The particle labelled in orange is now positioned to dominate the forces on the sevenfold particle, resulting in defect motion in the reverse direction. The schematic in Fig. \ref{defect_fbd} is corroborated by measurements of interaction forces over a driving cycle, shown in Fig. \ref{defect_ft}.

From this analysis, we see that the directionality imposed by the driving shear played an essential role in determining the defect motion. We have observed in Section \ref{protocol} that the driving shear causes a small anisotropy in the system's elastic response. The elasticity measurement was performed on a triangular lattice, where much of the driving cycle's effects are cancelled out by the symmetry of the lattice. The effect of anisotropy becomes more pronounced when defects are present as we saw above. Finally, this analysis shows that the defect motions in Fig. 4(a)-(c) is more aptly understood in terms of a core force $f^{\text{core}}$ rather from the Peach-Koehler force $f^{\text{PK}}$. The defect motion is sensitive to the microscopic details of bonds immediately neighbouring the defect, which is not contained in the coarse-grained continuum description contained in the Peach-Koehler force.

\subsection{Stability}
The grain boundary self-healing behaviour that we demonstrated in Fig. 4(a)-(c) of the main text required an important condition to be satisfied: the underlying crystal must itself be stable. Only in this case would the number of defects decrease over time as they self-propel and annihilate with one another. Otherwise, an unstable crystal would experience spontaneous generation of defects, and the self-healing we saw in Fig. 4(a)-(c) would not occur. In this section, we investigate the regimes of stability and instability in our system.\\

\begin{figure}[h!]
\centering
\includegraphics[width = .95\columnwidth]{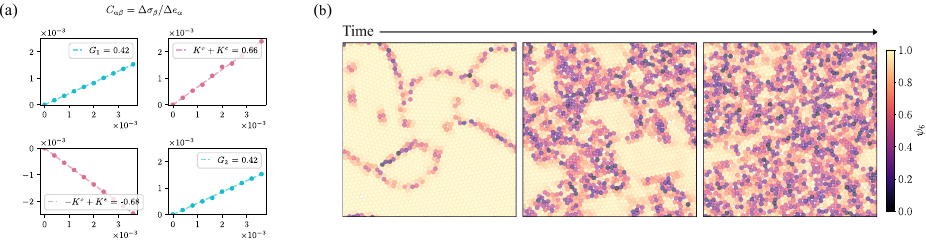}
\caption{A driven chiral system with friction coefficient $\mu = 4$ and chirality $\eps = 0.6$. (a) Shear and odd modulus measurements for the triangular lattice system. (b) A series of snapshots of the system (color representing $\psi_6$) shows that defects are spontaneously generated over time.}
\label{3-stability_mu4}
\end{figure}

In our system, the stability with respect to defect proliferation related to the relative strength of the tangential and normal forces; stronger tangential forces generally lead to greater instability. The effective interparticle force in main text Eq. (5) indicates that the ratio is approximately controlled by the quantity $\mu_- = \mu\,(1-\eps)$. In the results described in the main text, we focused on the case of $\mu = 1$, so that $\mu\,(1-\eps)$ was always smaller than 1. Let us now consider a system with $\mu = 4$ and chirality $\eps = 0.6$, with oscillatory driving shear as before. We note that $K^o \approx \mu\,(1-\eps)\, G = 1.6\,G$ for this system (Fig. \ref{3-stability_mu4}(a)). When this material is initialized with grain boundaries and allowed to evolve without external loading, it indeed experiences a proliferation of defects as depicted in the time series of Fig. \ref{3-stability_mu4}(b) (compare the series of snapshots in Fig. 3(a), where $\mu\,(1-\eps) = 0.4 < 1$).\\

\begin{figure}[h!]
\centering
\includegraphics[width = .95\columnwidth]{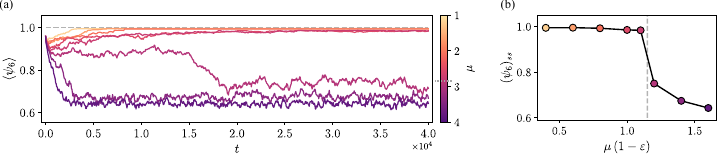}
\caption{Dependence of stability on the quantity $\mu\,(1-\eps)$ as $\mu$ is varied. (a) Time series of the average $\psi_6$. (b) The steady state value of $\langle \psi_6 \rangle$ vs the value of $\mu\,(1-\eps)$.}
\label{3-stability_v_mu}
\end{figure}

We examine the effect of the quantity $\mu\,(1-\eps)$ more systematically by performing simulations across a range of $\mu$ values between 1 (stable) and 4 (unstable), while keeping the chirality $\eps = 0.6$ fixed. In Fig. \ref{3-stability_v_mu}(a), we see that when $\mu\,(1-\eps)$ is small, the average $\psi_6$ of the system rapidly converges towards perfect crystalline order at $\psi_6 = 1$. In contrast, for larger $\mu\,(1-\eps)$ the average $\psi_6$ decreases to a value less than 1 and fluctuates thereafter close to this value. The plot of the final, steady state value of $\psi_6$ across the values of $\mu\,(1-\eps)$ shows that a transition between stable and unstable regimes occurs around $\mu\,(1-\eps) \approx 1.1$ for our system (Fig. \ref{3-stability_v_mu}(b)).\\

There is another distinct type of stability that is affected by the presence of transverse forces: the linear stability in terms of the spectrum of elastic waves. Previous works. Refs. \cite{scheibner_odd_2020, fruchart_odd_2023} have shown that an odd solid can experience oscillations even in an overdamped medium, and the condition for such ``active waves" is that $K^o(K^o-A) > (B/2)^2$. Noting that $A\approx 0$ and $B\approx 2G$ in our system, this corresponds to the condition that $K^o > G$. In Fig. 3, we demonstrated that $K^o$ can often be predicted by the formula $K^o = \mu\,(1-\eps)\, G$ in the triangular lattice of driven chiral particles. Because the relation of $K^o$ to $G$ mirrors the relation of transverse forces to normal forces $\abs{\mathbf{f}_t} \approx \mu\,(1-\eps) \abs{\mathbf{f}_n}$, in our system the regimes of linear stability roughly coincide with regimes where defects do not proliferate.\\

In summary, the analysis in this section shows that the evolution of the polycrystalline system is controlled by the quantity $\mu\,(1-\eps)$, which approximately describes the relative strength of the frictional and normal forces. When the ratio is small, the system undergoes spontaneous grain boundary motion that results in a decay towards crystalline order Fig. 3(a); as the ratio is increased, the system sustains a disordered state where defects are constantly generated and annihilated.

\vfill
\pagebreak
\section{Amorphous systems}
Here, we discuss the emergent spinning speeds, elasticity measurements of the amorphous system, and methods to characterize the chiral plastic vortices.

\subsection{Emergent $\Omega$ and odd moduli}
\label{4-measurements}

\begin{figure}[h!]
\centering
\includegraphics[width = .6\columnwidth]{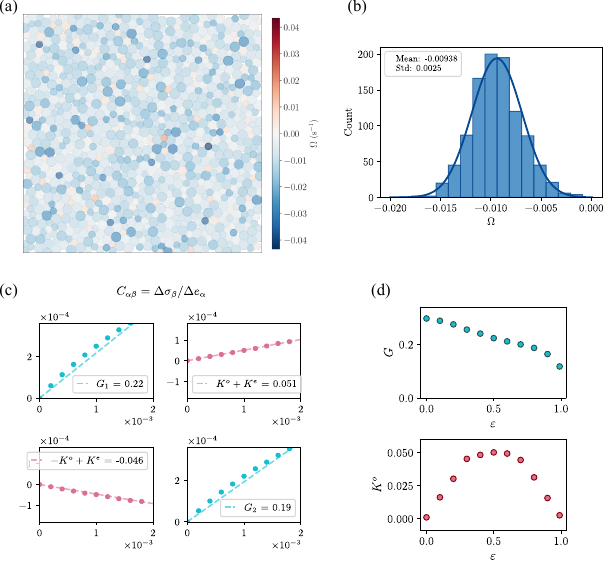}
\caption{Measurements of the amorphous system. (a) Map of emergent spinning speed $\Omega$. (b) Distribution of spinning speeds. (c) Shear and odd modulus measurements for the amorphous system (with chirality $\eps = 0.6$ and driving frequency $T_d = 1$). (d) Dependence of the moduli on the degree of chirality $\eps$. Compare the polycrystalline case shown in Fig. \ref{3-defect_C}.}
\label{4-amorphous-C}
\end{figure}

As in the previous systems, the amorphous system composed of chiral particles and driven by oscillatory shear shows emergent spinning and an odd elastic response. The map of spinning speeds, however, is strongly affected by the amorphous nature of this system. Whereas the polycrystalline system showed patches of uniform spinning, the amorphous system results in a highly disordered $\Omega$ field without large regions of uniformity (Fig \ref{4-amorphous-C}(a)). The spinning speed $\Omega$ of each particle is sensitive to the positions of its neighbours, so the amorphous structure of the solid becomes reflected in the $\Omega$ field. Nevertheless, the distribution of $\Omega$ is similar to the polycrystalline case with a spread around a clockwise average value (Fig \ref{4-amorphous-C}(b)). The elastic properties of the amorphous system is likewise similar to the polycrystalline case. We measure the elastic moduli using the same protocol as before, with a suitable choice of relaxation times considering the slow deformations shown in (Fig. 4(d)-(e)). Despite the highly disordered spatial distribution of spinning speeds, the system again shows an odd response and a non-monotonic dependence of $K^o$ on $\eps$ (Figs. \ref{4-amorphous-C}(c)-(d)).

\subsection{Characterization of vortices}

\begin{figure}[h!]
\centering
\includegraphics[width = .8\columnwidth]{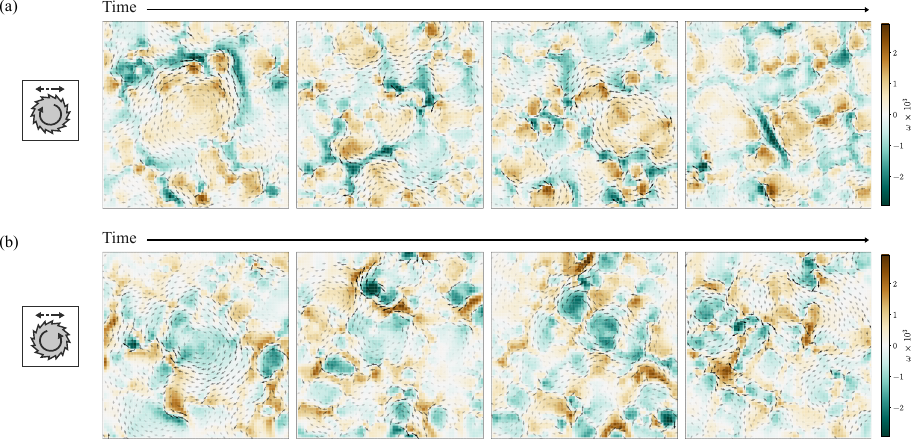}
\caption{Time evolution of the vorticity of the coarse-grained velocity field, $\omega = \boldsymbol{\nabla}\times \mb{v}$, for (a) clockwise spinning and (b) counter-clockwise spinning particles.}
\label{4-vortices_time}
\end{figure}

The slow, spontaneous deformations of the amorphous system exhibit vortical patches as shown in the series in Fig. \ref{4-vortices_time}, a snapshot of which is included in Fig. 4(e). Although the vortices are transient and evolve over time, they appear to have a consistent handedness that is determined by the chirality of the particles.\\ 

\begin{figure}[h!]
\centering
\includegraphics[width = .9\columnwidth]{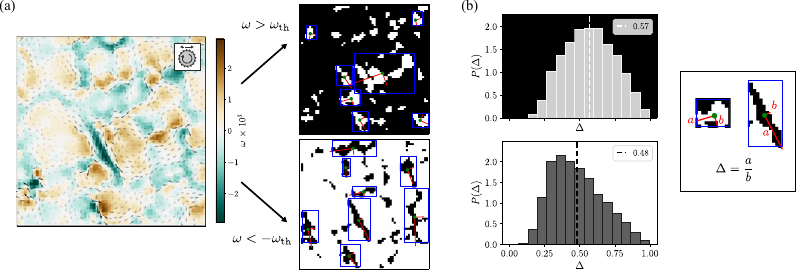}
\caption{Analysis of the vorticity field for the case of $\eps = 0.6$. (a) The $\omega$ field is binarized with a threshold $\omega_{\text{th}}$. (b) Distributions of the aspect ratio $\Delta = a/b$.}
\label{4-vortices_analysis}
\end{figure}

We now describe a method to quantify the preferred handedness of the chiral plastic vortices, and example of which is shown in Fig. \ref{4-vortices_analysis} for the case of $\eps>0$. For each snapshot of the vorticity field $\omega$, we binarize it using a threshold $\omega_{\text{th}}$: values with $\omega>\omega_{\text{th}}$ and values $\omega<-\omega_{\text{th}}$ are identified separately as clusters, shown in the binary plots in Fig. \ref{4-vortices_analysis}(a). We then compute the moment of inertia tensor for each cluster, and examine how the aspect ratio of the clusters $\Delta = a/b$ is distributed. The lengths $a$ and $b$ are the major and minor axes of the ellipse with the same second central moments of inertia tensor as the cluster being measured. The resulting distributions in Fig. \ref{4-vortices_analysis}(b) show that regions of positive vorticity $\omega>\omega_{\text{th}}$ have a larger aspect ratios than regions of negative vorticity $\omega<-\omega_{\text{th}}$. In other words, the deformations tend to form vortical patches with positive vorticity, surrounded by channels of negative vorticity between the different patches.\\

\begin{figure}[h!]
\centering
\includegraphics[width = .6\columnwidth]{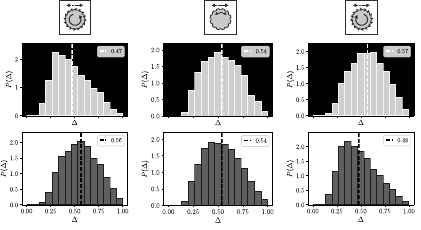}
\caption{Distributions of the aspect ratio $\Delta = a/b$ for (a) $\eps = -0.6$,  (b) $\eps = 0$, and (c) $\eps = 0.6$,.}
\label{4-vortices_compare}
\end{figure}

Finally, we compare the results between different values of $\eps$ to demonstrate the dependence of vorticity on the chirality. Fig. \ref{4-vortices_compare}(a) shows that when the chirality (and therefore the emergent spinning speed $\Omega$) is reversed, the deformations prefer to form patches of negative vorticity. In the null experiment with $\eps=0$ shown in Fig. \ref{4-vortices_compare}(b), patches with positive and negative vorticity form with equal likelihood.

\vfill\pagebreak

\bibliography{supp_references}